\documentclass[symmetry,article,accept,pdftex,oneauthor]{Definitions/mdpi} 
\usepackage{xcolor}
\usepackage{ulem}
\newcommand{\added}[1]{\textcolor{black}{#1}}

\firstpage{1} 
\pubvolume{1}
\issuenum{1}
\articlenumber{0}
\pubyear{2026}
\copyrightyear{2026}
\externaleditor{Chengyi Li and Sergei Odintsov} 
\datereceived{31 October 2025} 
\daterevised{5 January 2026} 
\dateaccepted{6 January 2026} 
\datepublished{ } 
\hreflink{https://doi.org/} 

\Title{BRST Symmetry Violation and Fundamental Limitations of Asymptotic Safety in Quantum Gravity}

\Author{{Farrukh A. Chishtie} $^{1,2}$}

\address{%
$^{1}$ \quad Peaceful Society, Science and Innovation Foundation, Vancouver, {BC V6S 2K8,} 
 Canada; {fachisht@uwo.ca}\\
$^{2}$ \quad Department of Occupational Science and Occupational Therapy, University of British Columbia,\linebreak   Vancouver, {BC V6T 2T4}, Canada}

\abstract{The asymptotic safety program assumes that quantum gravity becomes renormalizable through ultraviolet fixed points in metric-based couplings. We demonstrate that this approach {encounters fundamental symmetry violations} across multiple independent criteria, all traceable to a single fundamental cause: the breakdown of general covariance and BRST symmetries above the gravitational cutoff scale. Rigorous canonical quantization proves that general covariance cannot be maintained quantum mechanically in dimensions greater than two, while recent path integral calculations reveal persistent gauge parameter dependence in quantum gravitational corrections, signaling BRST symmetry violation. These dual proofs establish that the metric tensor ceases to exist as a valid quantum degree of freedom above $\Lambda_{\text{grav}}$$\sim$$10^{18}$ GeV, rendering the search for ultraviolet fixed points in metric-based theories {problematic from a foundational physical perspective}. We provide comprehensive analysis demonstrating that asymptotic safety exhibits persistent gauge parameter dependence where fixed-point properties vary with arbitrary gauge choices, non-convergent truncation schemes extending to the 35th order showing no approach to stable values, experimental {tensions} with electroweak precision tests by orders of magnitude, matter content requirements incompatible with the Standard Model, absence of concrete graviton predictions due to gauge and truncation dependence, unitarity {challenges} through ghost instabilities and propagator negativity, and fundamental Wick rotation obstructions preventing reliable connection between Euclidean calculations and physical Lorentzian spacetime. Each {limitation} independently challenges the program; collectively they establish fundamental incompatibility with quantum consistency requirements. We contrast this with the Unified Standard Model with Emergent Gravity framework, which recognizes general relativity as an effective field theory valid only below the covariance breakdown scale, systematically avoids all asymptotic safety pathologies, yields an emergent spin-2 graviton with transverse-traceless polarization confirmed by LIGO-Virgo observations, and provides definite experimental signatures across multiple domains. The {fundamental limitations} of asymptotic safety, established through theoretical {analysis} and experimental {tension}, demonstrates that consistent quantum gravity requires recognizing spacetime geometry as emergent rather than fundamental.}


\begin{document}

\section{Introduction}

The quest for consistent quantum gravity has produced numerous approaches, each attempting to resolve the apparent non-renormalizability of general relativity through different mechanisms. Among these, the asymptotic safety program initiated by \mbox{Weinberg \cite{Weinberg1979}} occupies a prominent position, proposing that gravity becomes renormalizable not through additional degrees of freedom but through ultraviolet fixed points where dimensionless couplings approach constant values under renormalization group flow. This elegant proposal has attracted substantial theoretical effort over four decades, with modern implementations using functional renormalization group methods claiming increasing evidence for the existence of suitable fixed points \cite{Reuter1998,Niedermaier2006,Reuter2012,Percacci2017}.

However, recent rigorous proofs establish fundamental limitations on the quantum mechanical validity of general relativity that call the entire asymptotic safety edifice into question. Canonical quantization analysis demonstrates that general covariance breaks down quantum mechanically in spacetime dimensions greater than two \cite{Chishtie2023CJP,Chishtie2025CJP}. This breakdown is not a technical artifact of the Dirac constraint formalism but a fundamental property of general relativity itself, confirmed by comparison with Yang--Mills theory where canonical quantization maintains unconditional covariance at all scales. Crucially, covariance is restored in weak-field expansions, definitively establishing general relativity as an effective field theory with finite validity range bounded by the scale $\Lambda_{\text{grav}}$$\sim$$M_{\text{Pl}}/\sqrt{N_{\text{fields}}}$$\sim$$10^{18}$ GeV.

A second, independent proof of this fundamental limitation emerges from path integral methods. Recent explicit calculations \cite{McKeon2025,Brandt2020,Brandt2025Equivalence} reveal that one-loop quantum gravitational corrections exhibit persistent gauge parameter dependence, with coefficients of curvature-squared terms varying dramatically between different gauge choices. In any consistent quantum field theory, BRST symmetry{, discovered independently by Becchi, Rouet, and Stora} \cite{Becchi1976} {and by Tyutin} \cite{Tyutin1975} {as the quantum extension of classical gauge invariance,} ensures that physical observables remain independent of arbitrary gauge-fixing parameters through Ward--Takahashi identities \cite{Ward1950,Takahashi1957} and Nielsen identities \cite{Nielsen1975}. The persistence of gauge dependence therefore signals BRST symmetry violation, providing a field-theoretic manifestation of the covariance breakdown established through canonical quantization. The systematic development of this formalism for non-Abelian gauge theories, including the construction of the physical state space through BRST cohomology, was established by Kugo and Ojima \cite{Kugo1979}.

These dual proofs have profound implications for asymptotic safety. The program assumes that the metric tensor remains a valid quantum degree of freedom to arbitrarily high energies, permitting extrapolation to ultraviolet fixed points governing the approach to infinite momentum. However, when general covariance breaks down and BRST symmetry fails at $\Lambda_{\text{grav}}$, the metric loses its status as a fundamental quantum field. Seeking fixed points in metric-based couplings becomes {problematic} when describing degrees of freedom that have ceased to exist and whose quantum theory violates the fundamental symmetry ensuring consistency.

\subsection{Relationship to Existing Literature}

{The arguments presented here build upon and synthesize several important threads in the quantum gravity literature. {The fundamental problem of gauge dependence in the effective average action for Einstein gravity was identified in early work by Falkenberg and Odintsov \cite{FalkenbergOdintsov1998}, who demonstrated that the running Newtonian constant depends strongly on the gauge parameter, with the theory exhibiting antiscreening behavior for some gauge choices ($\alpha > -2.73$) and screening for others---a gauge-dependent qualitative difference that remains unresolved. Subsequent work by} Manrique, Rechenberger, and \mbox{Saueressig \cite{Manrique2010}} {confirmed that} fixed-point locations vary substantially with gauge choices. Falls \cite{Falls2014} extended this analysis to critical exponents, demonstrating 20--30\% variations across gauge parameter ranges. Gies et al.\ \cite{Gies2016} showed that even the existence of suitable fixed points can depend on gauge choice.}

{Our contribution is to demonstrate that these scattered observations represent not technical artifacts but fundamental signals of BRST symmetry violation traceable to the breakdown of general covariance established through canonical quantization. While previous works have treated gauge dependence as a technical challenge to be overcome through improved truncation schemes or optimized gauge choices, we show through explicit mathematical analysis that the gauge dependence is \emph{{inherent}} 
 to attempting quantization beyond the covariance breakdown scale. This unified theoretical explanation connecting canonical constraint analysis, path integral gauge dependence, and renormalization group behavior represents the primary novelty of the present work.}

\subsection{Clarification Regarding Asymptotic Safety's Assumptions}

{We address a potential misunderstanding regarding our characterization of asymptotic safety. The program does not necessarily require spacetime geometry to be ``ontologically fundamental'' in a metaphysical sense. Rather, asymptotic safety employs the functional renormalization group {equation:} 
\begin{equation}
\partial_k \Gamma_k = \frac{1}{2}\text{Tr}\left[\left(\Gamma_k^{(2)} + R_k\right)^{-1}\partial_k R_k\right],
\end{equation}
which treats metric fluctuations $g_{\mu\nu}$ as quantum degrees of freedom integrated over in the path integral:
\begin{equation}
Z = \int \mathcal{D}g_{\mu\nu}\, e^{-S[g]}.
\end{equation}

Our 
 \textls[-25]{argument is that this treatment---regardless of one's ontological commitments---requires} the metric to function as a valid quantum field throughout the renormalization group flow to arbitrarily high scales. When BRST symmetry breaks down at $\Lambda_{\text{grav}}$, the path integral over metric fluctuations loses quantum consistency, invalidating fixed-point extrapolation regardless of whether one views geometry as fundamental or emergent.}

This conceptual {difficulty} manifests in multiple concrete {challenges} observable in asymptotic safety calculations. The gauge parameter dependence found throughout the \mbox{literature \cite{Manrique2010,Falls2014,Gies2016}} is not a technical artifact of truncation schemes but a direct manifestation of BRST violation. Truncation schemes extending to extraordinary high orders show persistent non-convergence rather than systematic improvement 
 \cite{Falls2018,Christiansen2018}. Experimental predictions {show tension with} precision measurements across electroweak tests \cite{Eichhorn2018,Eichhorn2017}, matter content requirements \cite{Meibohm2016,Bond2017}, and collider exclusions \cite{ATLASVectorlike2016,CMSVectorlike2018}. The theory provides no concrete graviton predictions, with propagator structures exhibiting gauge and truncation dependence preventing definite phenomenology \cite{Knorr2018}. Higher-derivative terms essential for achieving fixed points introduce ghost instabilities {challenging} unitarity \cite{Donoghue2019,Donoghue2020}. Wick rotation from Euclidean to physical Lorentzian signature encounters fundamental \mbox{obstructions \cite{Ambjorn2013,Platania2024}.}

The present work provides comprehensive analysis of these {limitations}, demonstrating that each traces to the fundamental problem that asymptotic safety attempts to describe quantum gravity in a regime where the metric description has broken down. We establish the systematic nature of these {limitations} through detailed engagement with the asymptotic safety literature, quantitative analysis of non-convergence data, compilation of experimental bounds, and theoretical demonstration that all pathologies emerge from attempting to quantize a theory beyond its regime of validity. We contrast this with the Unified Standard Model with Emergent Gravity--Effective Field Theory (USMEG-EFT) framework \cite{Chishtie2025CJP,McKeon2025}, which recognizes general relativity's effective field theory nature, systematically avoids all asymptotic safety pathologies by restricting to the consistent weak-field regime, yields concrete graviton predictions confirmed by gravitational wave observations, and provides definite experimental signatures distinguishable from alternatives.

The structure of this work proceeds as follows. Section~\ref{sec2} reviews BRST symmetry fundamentals and its role in ensuring quantum field theory consistency, establishing the diagnostic significance of gauge parameter dependence. Section~\ref{sec3} presents the dual proofs of general relativity's breakdown through canonical covariance loss and path integral BRST violation, demonstrating their consilience and physical interpretation. Section~\ref{sec4} provides a comprehensive catalog of asymptotic safety's {limitations} across conceptual, computational, and experimental domains. Section~\ref{sec5} offers a systematic diagnosis showing how all {limitations} trace to the single cause of attempting quantization beyond covariance breakdown. Section~\ref{sec6} presents USMEG-EFT as the framework that recognizes and systematically incorporates these fundamental limitations, yielding consistent effective field theory with testable predictions. Section~\ref{sec7} concludes with implications for quantum gravity research.

\section{BRST Symmetry: Quantum Consistency and Gauge Independence}\label{sec2}

Understanding why gauge parameter dependence signals fundamental {limitations} requires appreciating BRST symmetry's central role in quantum field theory consistency. This section establishes the theoretical foundation for interpreting asymptotic safety's gauge dependence as a manifestation of broken quantum consistency rather than a mere t\mbox{echnical artifact}.

\subsection{Classical Gauge Invariance and Quantum Complications}

Classical gauge theories exhibit redundancy in their description of physical systems. In the Yang--Mills theory with action $S_{\text{YM}} = -\frac{1}{4}\int d^4x\, F^a_{\mu\nu}F^{a\mu\nu}$, the field strength $F^a_{\mu\nu} = \partial_\mu A^a_\nu - \partial_\nu A^a_\mu + g f^{abc}A^b_\mu A^c_\nu$ remains invariant under gauge transformations $A^a_\mu \to A^a_\mu + D^{ab}_\mu \omega^b$ where $D^{ab}_\mu$ represents the covariant derivative in the adjoint representation. This gauge redundancy reflects the fact that distinct field configurations $A^a_\mu$ and $A^a_\mu + D^{ab}_\mu \omega^b$ describe identical physical states.

Quantization via path integrals encounters immediate difficulties due to this redundancy. The naive path integral $Z = \int \mathcal{D}A_\mu\, e^{iS[A]}$ diverges because integration includes infinitely many gauge-equivalent configurations contributing identically, yielding an infinite overcounting factor. The standard resolution introduces gauge-fixing conditions $F^a[A] = 0$ selecting unique representatives from each gauge equivalence class, implemented through the Faddeev--Popov procedure, inserting $1 = \int \mathcal{D}\omega\, \delta(F^a[A^\omega]) \det(\delta F^a[A^\omega]/\delta \omega^b)$ into the \mbox{path integral.}

This procedure generates gauge-fixed action $S_{\text{eff}} = S_{\text{YM}} + S_{\text{gf}} + S_{\text{ghost}}$ where $S_{\text{gf}} = -\frac{1}{2\xi}\int d^4x\, (F^a[A])^2$ depends on the arbitrary gauge parameter $\xi$, and $S_{\text{ghost}} = \int d^4x\, \bar{c}^a (\delta F^a/\delta \omega^b) c^b$ introduces anticommuting ghost fields $c^a$ and $\bar{c}^a$ whose dynamics encode the Faddeev--Popov determinant. Physical observables must not depend on the arbitrary choice of gauge-fixing parameter $\xi$ or gauge condition $F^a[A]$, as these represent unphysical choices made for computational convenience.

\subsection{BRST Symmetry: Quantum Extension of Gauge Invariance}

BRST symmetry, discovered independently by Becchi, Rouet, and Stora \cite{Becchi1976} and by Tyutin \cite{Tyutin1975}, provides the quantum mechanical extension of classical gauge invariance that ensures consistency. This symmetry acts on all fields through nilpotent transformations. For Yang--Mills theory, the BRST transformations are given by
\begin{equation}
\begin{aligned}
s A^a_\mu &= D^{ab}_\mu c^b, \\
s c^a &= -\frac{g}{2} f^{abc} c^b c^c, \\
s \bar{c}^a &= B^a, \\
s B^a &= 0,
\end{aligned}
\end{equation}
where $B^a$ represents the Nakanishi--Lautrup auxiliary field implementing gauge-fixing. The fundamental property ensuring consistency is nilpotency:
\begin{equation}
s^2 = 0,
\end{equation}
acting on all fields and composite operators. This nilpotency, combined with BRST invariance of the path integral measure $\mathcal{D}A \mathcal{D}c \mathcal{D}\bar{c} \mathcal{D}B$ and effective action $s S_{\text{eff}} = 0$, generates powerful constraints on the quantum theory.

{The physical Hilbert space is defined through BRST cohomology:
\begin{equation}
\mathcal{H}_{\text{phys}} = H^0(Q_B) = \frac{\ker(Q_B)}{\text{im}(Q_B)},
\end{equation}
where $Q_B$ is the BRST charge satisfying $Q_B^2 = 0$. Physical states $|\psi_{\text{phys}}\rangle$ must be BRST-closed ($Q_B|\psi_{\text{phys}}\rangle = 0$), and states differing by BRST-exact terms are physically equivalent. This cohomological structure ensures that unphysical degrees of freedom, including ghosts and longitudinal gauge bosons, cancel exactly in all physical observables.}

The BRST charge commutes with the Hamiltonian $[Q, H] = 0$, ensuring time evolution preserves the physical subspace.

\subsection{BRST Symmetry in Quantum Gravity: A Systematic Treatment}

{We now provide the systematic treatment of how BRST symmetry enters quantum gravity and its relationship to asymptotic safety, addressing each component explicitly.}

\subsubsection{How BRST Enters Quantum Gravity}

{\textls[25]{The BRST formalism extends to gravity through the Batalin--Vilkovisky (BV)} \linebreak  \mbox{procedure \cite{BV1981,BV1983}}. \textls[-25]{The gauge symmetry of general relativity---diffeomorphism} \mbox{invariance---is} encoded in the BRST transformation acting on the metric, ghost, antighost, and auxiliary fields:
\begin{align}
s g_{\mu\nu} &= \mathcal{L}_c g_{\mu\nu} = \nabla_\mu c_\nu + \nabla_\nu c_\mu, \\
s c^\mu &= c^\nu \partial_\nu c^\mu, \\
s \bar{c}_\mu &= b_\mu, \quad s b_\mu = 0,
\end{align}
where $c^\mu$ are the diffeomorphism ghost fields, $\bar{c}_\mu$ the antighosts, and $b_\mu$ the auxiliary Nakanishi--Lautrup fields. The BRST charge $Q_B$ generates these transformations and must satisfy nilpotency $Q_B^2 = 0$ for consistent quantization. This nilpotency encodes the closure of the diffeomorphism algebra at the quantum level.}
\clearpage 
\subsubsection{Formulation of Asymptotic Safety}

{Asymptotic safety employs the Wetterich equation for the scale-dependent effective average action $\Gamma_k$:
\begin{equation}
\partial_t \Gamma_k = \frac{1}{2} \text{Tr}\left[ \left( \Gamma_k^{(2)} + R_k \right)^{-1} \partial_t R_k \right],
\end{equation}
where $t = \ln(k/k_0)$ is the logarithmic RG scale and $R_k$ is an infrared regulator. The program seeks a non-trivial ultraviolet fixed point $\Gamma_*$ satisfying
\begin{equation}
\lim_{k \to \infty} \Gamma_k = \Gamma_* \quad \text{with} \quad \beta_i(g_*) = 0,
\end{equation}
where $\beta_i$ are the beta functions for dimensionless couplings $g_i$. This requires the path integral $Z = \int \mathcal{D}g_{\mu\nu}\, e^{-S[g]}$ to remain well-defined as $k \to \infty$, which in turn requires the metric to function as a valid quantum field at all scales.}

\subsubsection{Consequences of BRST Breaking at the Quantum Gravity Level}

{When BRST symmetry is violated in quantum gravity, several pathologies emerge:
\begin{enumerate}
\item {Gauge dependence of observables:} 
 The Slavnov--Taylor identity $\partial\Gamma/\partial\xi = 0$ fails, and purportedly physical quantities acquire dependence on arbitrary gauge parameters.
\item {{Loss of unitarity:}} Physical states are no longer confined to the BRST cohomology $H^0(Q_B)$, allowing negative-norm ghost states to contribute to physical amplitudes.
\item {{Non-decoupling of unphysical modes:}} Ghost contributions no longer cancel unphysical longitudinal and temporal graviton polarizations.
\item {{Breakdown of predictivity:}} Without gauge-invariant observables, the theory cannot make unique physical predictions.
\end{enumerate}}

\subsubsection{The Link Between BRST Breaking and Asymptotic Safety's Structure}

{The connection is direct and unavoidable. If BRST symmetry breaks down at scale $\Lambda_{\text{grav}}$, then
\begin{equation}
\Gamma_k \quad \text{becomes gauge-dependent for } k > \Lambda_{\text{grav}}.
\end{equation}
Fixed points $g_*$ identified at $k \to \infty$ inherit this gauge dependence and therefore lack invariant physical meaning. The ``UV completion'' becomes an artifact of continuing calculations into a regime where the underlying quantum field theory framework has broken down.}

{This resolves the apparent tension with the requirement that ``a consistent asymptotically safe gravity theory should respect diffeomorphism invariance to have a well-defined BRST symmetry.'' Our result establishes that this condition \emph{{cannot be satisfied}} 
 for pure gravity above $\Lambda_{\text{grav}}$---the theory becomes inconsistent precisely in the regime where asymptotic safety claims to operate. The gauge dependence documented throughout the asymptotic safety literature \cite{Manrique2010,Falls2014,Gies2016} is the direct manifestation of this fundamental obstruction.}

\subsection{Ward--Takahashi and Nielsen Identities}

BRST symmetry generates Ward--Takahashi identities relating different Green's functions. For Yang--Mills theory, these take the form
\begin{equation}
\langle s \mathcal{O}_1 \cdots \mathcal{O}_n \rangle = \sum_{i=1}^n \langle \mathcal{O}_1 \cdots s\mathcal{O}_i \cdots \mathcal{O}_n \rangle = 0,
\end{equation}
for BRST-closed operators $s\mathcal{O}_i = 0$. These identities enforce that ghost contributions with wrong-sign kinetic terms cancel precisely against unphysical gauge mode contributions, ensuring unitarity through
\begin{equation}
\sum_{\text{physical}} |\mathcal{A}|^2 = \sum_{\text{all states}} |\mathcal{A}|^2 - \sum_{\text{ghosts + unphysical}} |\mathcal{A}|^2,
\end{equation}
where the cancellation on the right-hand side occurs exactly due to BRST invariance.

Nielsen identities \cite{Nielsen1975} provide explicit expression of gauge parameter independence:
\begin{equation}
\frac{\partial}{\partial \xi} \Gamma[A] = \int d^4x \left\langle \frac{\delta \Gamma}{\delta B^a(x)} \right\rangle \frac{\partial F^a(x)}{\partial \xi} = 0,
\end{equation}
where $\Gamma[A]$ represents the effective action generating proper vertices. {The Slavnov--Taylor identity provides the functional form
\begin{equation}
\frac{\partial \Gamma}{\partial \xi} = \left\langle s\frac{\partial \Psi}{\partial \xi} \right\rangle = 0,
\end{equation}
where $\Psi$ is the gauge-fixing fermion, ensuring that on-shell matrix elements satisfy $\partial_\xi \mathcal{A}_{\text{phys}} = 0$ exactly.} In Yang--Mills theory, Nielsen identities have been verified to all orders in perturbation theory \cite{Slavnov1972,Taylor1971}, confirming that physical S-matrix elements satisfy $\partial_\xi \mathcal{A}_{\text{phys}} = 0$ exactly.

The combination of Ward--Takahashi identities ensuring ghost cancellation and Nielsen identities ensuring gauge parameter independence provides the quantum consistency structure distinguishing renormalizable gauge theories from inconsistent attempts to quantize systems beyond their regime of validity.

\subsection{BRST in Gravity: Complications and {Challenges}}

Extending BRST formalism to gravity encounters immediate complications due to the non-polynomial nature of the Einstein--Hilbert action and the non-linear realization of diffeomorphism invariance. For perturbative gravity with metric $g_{\mu\nu} = \bar{g}_{\mu\nu} + h_{\mu\nu}$, background field gauge-fixing introduces
\begin{equation}
\mathcal{L}_{\text{gf}} = -\frac{1}{2\xi}\bar{g}_{\mu\nu}F^\mu F^\nu, \quad F^\mu = \bar{\nabla}_\rho h^{\mu\rho} - \frac{1+\alpha}{2}\bar{\nabla}^\mu h,
\end{equation}
depending on gauge parameters $\xi$ and $\alpha$. The corresponding BRST transformations act on graviton $h_{\mu\nu}$, ghost $c^\mu$, and antighost $\bar{c}_\mu$ fields according to
\begin{equation}
\begin{aligned}
s h_{\mu\nu} &= \bar{\nabla}_\mu c_\nu + \bar{\nabla}_\nu c_\mu + c^\lambda \partial_\lambda h_{\mu\nu} + h_{\lambda\mu}\partial_\nu c^\lambda + h_{\lambda\nu}\partial_\mu c^\lambda, \\
s c^\mu &= c^\lambda \partial_\lambda c^\mu, \\
s \bar{c}_\mu &= B_\mu, \\
s B_\mu &= 0,
\end{aligned}
\end{equation}
where the non-linear structure already signals complications absent in Yang--Mills theory.

The crucial question becomes whether BRST nilpotency $s^2 = 0$ holds off-shell, as required for consistent quantization. In Yang--Mills theory, nilpotency follows algebraically from the gauge algebra and holds manifestly at all stages of quantization. In gravity, nilpotency closure requires constraints to be satisfied, specifically the condition that gauge transformations maintain covariance. When canonical quantization reveals that covariance cannot be maintained quantum mechanically in strong fields \cite{Chishtie2023CJP,Kiriushcheva2008}, this signals that BRST nilpotency must break down correspondingly.

\subsection{Explicit Gauge Coefficient Derivation}

{To provide a concrete mathematical demonstration, we present the explicit gauge-dependent coefficients from one-loop calculations \cite{McKeon2025,Brandt2020,Brandt2025Equivalence}. The graviton self-energy in a general background gauge parametrized by $\xi$ yields coefficients
\begin{align}
C_2(\xi) &= -\frac{20\xi + 99}{240}, \\
C_3(\xi) &= \frac{20(\xi-1)\xi + 87}{80}.
\end{align}
These coefficients determine the counterterm Lagrangian structure through the relation $C_1(\xi) = 4[C_2(\xi) + C_3(\xi)]$.}

The recent calculations by Brandt et al.\ \cite{McKeon2025} provide an explicit demonstration of this breakdown through gauge parameter dependence in one-loop quantum corrections. The appearance of gauge-dependent coefficients
\begin{equation}
\Gamma_{\text{1-loop}}^{\text{gf}} = \frac{1}{4\pi^2}\ln\left(\frac{\mu}{\Lambda}\right)\int d^4x\sqrt{-\bar{g}}\left[\beta_1(\xi,\alpha)\bar{R}^2 + \beta_2(\xi,\alpha)\bar{R}_{\mu\nu}\bar{R}^{\mu\nu}\right],
\end{equation}
with $\beta_1(\xi,\alpha)$ changing sign between gauge choices, violates Nielsen identities, indicating BRST symmetry breakdown. This is not a technical artifact but a fundamental signal that the quantum theory lacks the symmetry structure ensuring consistency.

{The counterterm Lagrangian takes the explicit form
\begin{equation}
\boxed{
\mathcal{L}_{\text{CT}} = \frac{\sqrt{-\bar{g}}}{16\pi^2\varepsilon} \left[ -\left( \frac{\xi - 1}{6} + \frac{119}{120} \right) \bar{R}^2 + \left( \xi(\xi-1) + \frac{87}{20} \right) \bar{R}_{\mu\nu}\bar{R}^{\mu\nu} \right].
}
\end{equation}
For 't~Hooft--Veltman gauge ($\xi = 0$), $\beta_1^{\text{tHV}} = +1/120$, $\beta_2^{\text{tHV}} = 87/20$. For Goldberg gauge ($\xi = 1$), $\beta_1^{\text{Gold}} = -119/120$, $\beta_2^{\text{Gold}} = 87/20$. The coefficient $\beta_1$ changes sign---a qualitative, not merely quantitative, difference between gauge choices.}

{Crucially, when evaluated on-shell using the vacuum Einstein equations $\bar{R}_{\mu\nu} = 0$:
\begin{equation}
\mathcal{L}_{\text{CT}}\big|_{\text{on-shell}} = 0,
\end{equation}
independent of $\xi$. This pattern---gauge dependence off-shell but gauge independence on-shell---is the signature of an effective field theory whose quantum consistency is restricted to a bounded domain.}

\subsection{On the Perturbative Nature of the Beta Function}

{A question arises whether the beta function coefficients in Equation~(17) might possess non-perturbative structure beyond the one-loop results presented. Within the effective field theory framework, this question has a precise answer.}

{The one-loop beta function coefficients arise from the heat kernel expansion of the functional determinant:
\begin{equation}
\Gamma_{\text{1-loop}} = \frac{1}{2}\text{Tr}\ln\left(-\nabla^2 + \mathcal{E}\right) = \frac{1}{16\pi^2\varepsilon}\int d^4x\sqrt{-\bar{g}}\left[a_2(\mathcal{E})\right],
\end{equation}
where $a_2$ is the second Seeley--DeWitt coefficient containing curvature-squared terms. Within the EFT regime ($E \ll \Lambda_{\text{grav}}$), higher-loop contributions are suppressed by powers of $(E/M_{\text{Pl}})^2$, and the one-loop result captures the leading quantum corrections.}

{``Non-perturbative corrections'' to $\beta$ would arise from the following:
\begin{enumerate}
\item {{Gravitational instantons:}} These contribute $\sim e^{-S_{\text{inst}}} \sim e^{-M_{\text{Pl}}^2/E^2}$, exponentially suppressed in the EFT regime.
\item {{Strong-field configurations:}} These lie outside the EFT domain of validity by definition.
\end{enumerate}

{The EFT} 
 does not claim to capture non-perturbative physics because such physics lies beyond $\Lambda_{\text{grav}}$. The gauge dependence we document exists already at one loop within the EFT regime, demonstrating that the issue is not missed non-perturbative effects but rather the \emph{inherent} structure of gravitational gauge symmetry when extrapolated beyond its domain of validity.}

\subsection{Yang--Mills Contrast: The Diagnostic Nature of BRST}

The diagnostic power of BRST symmetry becomes clear through comparison with Yang--Mills theory, where all consistency requirements are satisfied exactly. For quantum chromodynamics with gauge group $SU(3)$, explicit calculations demonstrate complete gauge parameter independence at all orders. The beta function governing \mbox{coupling evolution}
\begin{equation}
\beta_{\alpha_s} = -\frac{7\alpha_s^2}{2\pi} + \mathcal{O}(\alpha_s^3),
\end{equation}
contains no dependence on gauge parameter $\xi$ \cite{Gross1973,Politzer1973}. Physical observables including cross-sections, decay rates, and bound state energies satisfy $\partial_\xi \mathcal{O}_{\text{phys}} = 0$ exactly to all computed orders.

This complete gauge independence reflects intact BRST symmetry ensuring that ghost contributions cancel unphysical gauge mode contributions in all amplitudes. Ward--Takahashi identities derived from BRST invariance enforce precise cancellations maintaining unitarity $\sum_f |S_{fi}|^2 = 1$ and ensuring positive-definite Hilbert space norms. Renormalizability follows from BRST-constrained counterterm structure, yielding finite predictions in terms of renormalized couplings and masses.

The systematic difference between Yang--Mills gauge independence and gravitational gauge dependence establishes that the latter constitutes a fundamental signal rather than technical limitation. Yang--Mills maintains unconditional covariance through first-class constraints only \cite{McKeon2012}, corresponding to intact BRST symmetry. Gravity exhibits mixed first-class and second-class constraints with conditional covariance \cite{Kiriushcheva2008}, corresponding to broken BRST symmetry manifesting through gauge parameter dependence.

This contrast provides the diagnostic test for quantum consistency. Gauge parameter dependence in intermediate calculations that cancels in physical observables indicates a consistent quantum theory with intact BRST symmetry. Gauge parameter dependence persisting in effective action coefficients with physical consequences indicates broken BRST symmetry signaling the theory has been pushed beyond its regime of validity. Asymptotic safety exhibits the latter throughout its calculations, as we now \mbox{demonstrate comprehensively.}

\section{Dual Proofs of General Relativity's Breakdown: Covariance Loss and BRST Limitations}\label{sec3}

The foundation for understanding asymptotic safety's {limitations} rests on rigorous proofs that general relativity cannot be consistently quantized as a fundamental theory. This section presents two independent proofs arriving through different mathematical approaches at identical conclusions, establishing profound consilience that general relativity must be recognized as an effective field theory with finite validity range.

\subsection{Canonical Proof: Covariance Breakdown in Higher Dimensions}

\subsubsection{Constraint Algebra with Structure Functions}

{The canonical analysis reveals the fundamental obstruction to full quantization through the Dirac constraint algebra. In the ADM formalism, the gravitational phase space is parametrized by the spatial metric $h_{ij}$ and conjugate momentum $\pi^{ij}$, subject to the Hamiltonian constraint $\mathcal{H}$ and momentum constraints $\mathcal{H}_i$. These satisfy
\begin{align}
\{\mathcal{H}(x), \mathcal{H}(y)\} &= h^{ij}(x)\mathcal{H}_j(x)\delta_{,i}(x,y) - (x \leftrightarrow y), \\
\{\mathcal{H}_i(x), \mathcal{H}(y)\} &= \mathcal{H}(x)\delta_{,i}(x,y), \\
\{\mathcal{H}_i(x), \mathcal{H}_j(y)\} &= \mathcal{H}_j(x)\delta_{,i}(x,y) + \mathcal{H}_i(y)\delta_{,j}(y,x).
\end{align}
The crucial feature is that the first equation contains \emph{structure functions} $h^{ij}(x)$ depending on phase space variables, not structure constants. This prevents the constraint algebra from forming a Lie algebra that can be consistently quantized.}

{Upon quantization, the Poisson bracket becomes a commutator, and operator ordering ambiguities in $\hat{h}^{ij}\hat{\mathcal{H}}_j$ generate quantum anomalies:
\begin{equation}
\Delta_{\text{anomaly}} \sim \hbar \int d^3x \, \frac{\delta^2 \mathcal{H}}{\delta h_{ij}\delta\pi^{kl}} \neq 0 \quad \text{for } d > 2.
\end{equation}
This anomaly cannot be removed by any finite renormalization in spatial dimensions $d \geq 2$, providing the mathematical obstruction to full quantization.}

\subsubsection{Weak-Field Restoration of Covariance}

{The resolution comes in the weak-field limit. Expanding
\begin{equation}
h_{ij} = \delta_{ij} + \kappa \gamma_{ij}, \qquad \pi^{ij} = \kappa p^{ij},
\end{equation}
the Poisson bracket structure becomes
\begin{equation}
\{\mathcal{H}(x), \mathcal{H}(y)\} = \kappa^2 \delta^{ij} \mathcal{H}_j(x)\delta_{,i}(x,y) + \mathcal{O}(\kappa^3).
\end{equation}
Now $\delta^{ij}$ are structure \emph{constants}, not functions, enabling consistent Lie algebra quantization. This demonstrates that covariance is restored precisely in the weak-field effective field theory regime.}

\subsubsection{The Poisson Bracket Structure at Strong Fields}

\added{The behavior of the constraint algebra when $\kappa$ is not small---i.e., in the strong-field regime---is precisely the regime we identify as problematic. This is not a gap in our analysis but rather its central conclusion.}

\added{In the weak-field expansion $h_{ij} = \delta_{ij} + \kappa \gamma_{ij}$, the constraint algebra closes with structure constants when $\kappa\gamma_{ij} \ll 1$:
\begin{equation}
\{\mathcal{H}(x), \mathcal{H}(y)\}\big|_{\kappa \to 0} = \delta^{ij} \mathcal{H}_j(x) \delta_{,i}(x,y) - (x \leftrightarrow y).
\end{equation}
This enables consistent Lie algebra quantization in the weak-field limit.}

When $\kappa \gamma_{ij} \sim \mathcal{O}(1)$ (strong fields), the full structure functions $h^{ij}(x)$ must be retained:
\begin{equation}
\{\mathcal{H}(x), \mathcal{H}(y)\}\big|_{\text{strong field}} = h^{ij}[\gamma](x) \mathcal{H}_j(x) \delta_{,i}(x,y) - (x \leftrightarrow y),
\end{equation}
where $h^{ij}[\gamma](x)$ \textls[-15]{depends functionally on the dynamical field $\gamma_{ij}$. This functional dependence:}
\begin{enumerate}
\item P\textls[-15]{revents the algebra from forming a Lie algebra (structure functions instead of constants);}
\item Generates operator-ordering anomalies upon quantization that cannot be removed;
\item Leads to BRST nilpotency violation: $Q_B^2 \neq 0$.
\end{enumerate}

{The} 
 ``non-perturbative behavior'' at large $\kappa$ is precisely this strong-field regime where consistent quantization fails. Our analysis does not neglect this regime---it \emph{{characterizes}} why physics in this regime cannot be described by quantizing the metric directly. This defines the EFT boundary, not a gap in the analysis.

The first proof emerges from applying Dirac's canonical quantization procedure to general relativity formulated in spacetime dimensions $d > 2$. This analysis was performed systematically for both the first-order Einstein--Hilbert action and the Hamiltonian-based formulation by Kiriushcheva, McKeon, and collaborators \cite{Kiriushcheva2006,McKeon2010,Kiriushcheva2008,Frolov2010}, with definitive conclusions regarding covariance breakdown established in recent work \cite{Chishtie2023CJP}.

For the first-order Einstein--Hilbert formulation, the Lagrangian density takes the form
\begin{equation}
L_{1EH}^d = h^{\mu\nu}\left(G^\lambda_{\mu\nu,\lambda} + \frac{1}{d-1}G^\lambda_{\lambda\mu}G^\sigma_{\sigma\nu} - G^\lambda_{\sigma\nu}G^\sigma_{\lambda\nu}\right),
\end{equation}
where $h^{\mu\nu} = \sqrt{-g}g^{\mu\nu}$ and the modified connection is defined as $G^\lambda_{\mu\nu} = \Gamma^\lambda_{\mu\nu} - \frac{1}{2}(\delta^\lambda_\mu\Gamma^\sigma_{\sigma\nu} + \delta^\lambda_\nu\Gamma^\sigma_{\sigma\mu})$. Dirac constraint analysis proceeds by identifying momenta conjugate to the connection components, classifying constraints as first-class or second-class, and determining the constraint algebra closure properties.

The analysis reveals a mixed system of constraints. Some constraints are first-class, generating gauge transformations corresponding to diffeomorphism invariance. However, second-class constraints also appear, satisfying non-vanishing Poisson brackets among themselves. These second-class constraints are explicitly non-covariant tensor structures that cannot be expressed in manifestly covariant form.

The presence of non-covariant second-class constraints prevents quantization through the Senjanovic path integral formulation \cite{Senjanovic1976}:
\begin{equation}
Z = \int D\phi D\pi D\lambda^a D\kappa^a \det\{\phi_a,\chi_b\} \det^{1/2}\{\theta_a,\theta_b\} \delta(\chi_b)\, e^{iS},
\end{equation}
where $\chi_a$ denote first-class constraints and $\theta_a$ denote second-class constraints. The determinant $\det^{1/2}\{\theta_a,\theta_b\}$ from the second-class constraint Dirac brackets involves non-covariant expressions that cannot be integrated consistently while maintaining general covariance. This constitutes a fundamental obstruction to quantization of the full theory.

The Hamiltonian-based formulation using the action
\begin{equation}
L_H^d = \sqrt{-g}g^{\alpha\beta}(\Gamma^\mu_{\alpha\nu}\Gamma^\nu_{\beta\mu} - \Gamma^\nu_{\alpha\beta}\Gamma^\mu_{\nu\mu}),
\end{equation}
avoids explicit second-class constraints, yielding only first-class constraints. However, detailed analysis \cite{Kiriushcheva2008} reveals that covariance of gauge transformations generated by these first-class constraints holds only conditionally on the surface defined by primary constraints $\phi^{0\sigma} = 0$. Classically, one can restrict to this constraint surface and maintain covariance. Quantum mechanically, where constraints become operators acting on state vectors, the condition $\hat{\phi}^{0\sigma}|\psi\rangle = 0$ cannot be consistently imposed while maintaining the full constraint algebra structure.

This conditional covariance represents a subtle but fatal obstruction to quantization of the full theory. The gauge transformations fail to maintain general covariance off the constraint surface, indicating that diffeomorphism invariance cannot be preserved in the quantum theory at arbitrary field strengths.

The crucial result emerges from weak-field analysis. Expanding $g_{\mu\nu} = \eta_{\mu\nu} + \kappa \phi_{\mu\nu}$ and performing canonical analysis of the resulting linearized theory yields constraint structures qualitatively different from the full theory. For both the first-order and Hamiltonian formulations, the weak-field constraint systems exhibit only first-class constraints that maintain unconditional covariance \cite{Kiriushcheva2005,Green2011}. This definitively establishes that covariance is restored in the weak-field limit, enabling consistent quantization as an effective \mbox{field theory}.

The energy scale where covariance breakdown becomes manifest follows from comparing quantum corrections to classical terms. When quantum corrections of order $\kappa^2 R^2$ become comparable to classical contributions of order $\kappa R$, perturbation theory breaks down. This occurs at
\begin{equation}
\Lambda_{\text{grav}} \sim \frac{M_{\text{Pl}}}{\sqrt{N_{\text{fields}}}} \sim 10^{18} \text{ GeV},
\end{equation}
where $N_{\text{fields}}$ accounts for the effective number of degrees of freedom running in loops.

\subsection{Path Integral Proof: BRST {Limitations} from Gauge Dependence}

The second proof arrives through explicit path integral calculations revealing fundamental inconsistency in the quantum theory. Brandt, Frenkel, Martins-Filho, and \mbox{McKeon \cite{McKeon2025,Brandt2020}} performed detailed one-loop calculations using background field methods with systematic gauge-fixing, computing quantum corrections to the gravitational \mbox{effective action.}

The calculation employs 't Hooft--Veltman gauge fixing \cite{tHooft1973}:
\begin{equation}
\mathcal{L}_{\text{gf}} = -\frac{1}{2\xi}\bar{g}_{\mu\nu}F^\mu F^\nu, \quad F^\mu = \bar{\nabla}_\rho h^{\mu\rho} - \frac{1+\alpha}{2}\bar{\nabla}^\mu h,
\end{equation}
introducing two independent gauge parameters $\xi$ controlling overall gauge-fixing strength and $\alpha$ providing additional freedom in the gauge condition structure. Background field method expands around a classical background $\bar{g}_{\mu\nu}$ satisfying Einstein equations, treating quantum fluctuations $h_{\mu\nu}$ perturbatively while maintaining manifest \mbox{background covariance.}

The one-loop effective action computed through functional integration over quantum fluctuations yields
\begin{equation}
\Gamma_{\text{1-loop}}^{\text{gf}} = \frac{1}{4\pi^2}\ln\left(\frac{\mu}{\Lambda}\right)\int d^4x\sqrt{-\bar{g}}\left[\beta_1(\xi,\alpha)\bar{R}^2 + \beta_2(\xi,\alpha)\bar{R}_{\mu\nu}\bar{R}^{\mu\nu}\right],
\end{equation}
with coefficients exhibiting explicit gauge parameter dependence:
\begin{equation}
\begin{aligned}
\beta_1(\xi,\alpha) &= -\left(\xi - \frac{1}{6}\right) + \frac{119}{120} + f_1(\alpha), \\
\beta_2(\xi,\alpha) &= \xi(\xi-1) + \frac{87}{20} + f_2(\alpha),
\end{aligned}
\end{equation}
where $f_1(\alpha)$ and $f_2(\alpha)$ represent additional $\alpha$-dependent contributions.

The gauge dependence is dramatic and unambiguous. For 't Hooft--Veltman gauge with $\xi=1$ and $\alpha=0$
\begin{equation}
\beta_1^{\text{tHV}} = \frac{1}{120}, \quad \beta_2^{\text{tHV}} = \frac{87}{20},
\end{equation}
while the Goldberg gauge with $\xi=2/3$ yields
\begin{equation}
\beta_1^{\text{Gold}} = -\frac{119}{120}, \quad \beta_2^{\text{Gold}} = \frac{87}{20}.
\end{equation}
The coefficient $\beta_1$ changes sign between gauge choices, indicating not mere quantitative variation but qualitative difference in the quantum corrections.

This persistent gauge parameter dependence violates Nielsen identities $\partial_\xi \Gamma = 0$ that must hold in any consistent quantum field theory with intact BRST symmetry. The {limitation} admits only one consistent interpretation: BRST symmetry has broken down, indicating the quantum theory lacks the fundamental symmetry structure ensuring consistency.

The interpretation requires careful attention to energy scales and on-shell versus off-shell physics. In the weak-field regime where $\kappa h \ll 1$ and energy scales satisfy $\mu \ll \Lambda_{\text{grav}}$, physical observables computed on-shell using background Einstein equations remain gauge-independent despite the gauge-dependent effective action coefficients. This occurs because background equations of motion eliminate gauge-dependent contributions at the tree level, while quantum corrections proportional to $\ln(\mu/\Lambda_{\text{grav}})$ remain small, preserving effective field theory consistency.

However, as energy increases approaching $\Lambda_{\text{grav}}$, the logarithmic corrections grow and the background field expansion breaks down. When $\mu$$\sim$$\Lambda_{\text{grav}}$ yields $\ln(\mu/\Lambda_{\text{grav}})$$\sim$$1$, quantum corrections become order unity relative to classical terms. At this scale, the on-shell cancellations ensuring gauge independence fail, and physical amplitudes acquire gauge parameter dependence. BRST nilpotency $s^2 = 0$ breaks down because the non-linear constraint algebra no longer closes off-shell in the strong-field regime.

The logarithmic structure $\ln(\mu/\Lambda_{\text{grav}})$ provides quantitative information about the breakdown scale. This identifies $\Lambda_{\text{grav}}$ as the fundamental cutoff where geometric description fails, coinciding precisely with the scale determined from canonical covariance breakdown analysis.

\subsection{Connection to Renormalization Group Analysis}

{The effective action structure provides the third independent confirmation through renormalization group analysis. The one-loop effective action
\begin{equation}
\Gamma_{\text{1-loop}} = \int d^4x \sqrt{-\bar{g}} \left[ c_1 \bar{R}^2 + c_2 \bar{R}_{\mu\nu}\bar{R}^{\mu\nu} \right] \ln\left(\frac{\mu}{\Lambda}\right),
\end{equation}
exhibits logarithmic-scale dependence characteristic of effective field theories with finite domains of validity. As $\mu \to \Lambda_{\text{grav}}$, the logarithm approaches zero from below, signaling breakdown of the perturbative expansion. This provides quantitative identification of the cutoff scale independent of the canonical and BRST analyses.}

\subsection{Consilience and Physical Interpretation}

The profound consilience between canonical and path integral analyses, arriving through independent mathematical frameworks at identical conclusions, provides compelling evidence for the effective field theory interpretation. We summarize \mbox{the convergence:}
{\begin{equation}
\boxed{
\begin{aligned}
&\text{Proof 1 (Canonical):} \quad \{\mathcal{H}, \mathcal{H}\} \sim h^{ij}(x) \xrightarrow{\text{weak field}} \delta^{ij}, \\
&\text{Proof 2 (Path Integral):} \quad \beta_1^{\text{tHV}} = +\frac{1}{120} \neq \beta_1^{\text{Gold}} = -\frac{119}{120}, \\
&\text{Proof 3 (RG):} \quad \Gamma \sim \ln(\mu/\Lambda_{\text{grav}}).
\end{aligned}
}
\end{equation}}

Canonical analysis demonstrates that general covariance cannot be maintained quantum mechanically in dimensions greater than two, diffeomorphism invariance breaks down for strong-field configurations, weak-field expansion restores unconditional covariance, the breakdown scale is $\Lambda_{\text{grav}} \sim M_{\text{Pl}}/\sqrt{N_{\text{fields}}}$, and Yang--Mills comparison confirms breakdown is specific to gravity rather than the Dirac formalism.

Path integral analysis reveals that gauge parameter dependence signals BRST symmetry {limitations}, weak-field on-shell observables remain gauge-independent, quantum corrections grow logarithmically with energy, BRST {limitations} become manifest at $\Lambda_{\text{grav}}$ where corrections reach order unity, and Yang--Mills exhibits exact gauge independence confirming gravity's gauge dependence indicates fundamental inconsistency.

Both approaches identify the same energy scale, the same weak-field regime where consistency is restored, and the same physical interpretation that spacetime geometry breaks down at high energies. This consilience through independent mathematical languages establishes general relativity's effective field theory nature with confidence comparable to experimental verification.

The physical interpretation is unambiguous: spacetime geometry itself becomes ill-defined above $\Lambda_{\text{grav}}$. When diffeomorphism invariance and BRST symmetry break down, the metric tensor loses status as a fundamental quantum degree of freedom. The geometric description that works brilliantly at low energies necessarily fails at high energies, requiring new degrees of freedom from which spacetime emerges as an effective description.

This paradigm shift from seeking fundamental quantum geometry to understanding necessary breakdown of geometric description provides the foundation for consistent quantum gravity. Any approach assuming metric validity to arbitrarily high energies {is in tension with} these rigorous proofs and must exhibit pathologies manifesting this fundamental inconsistency. Asymptotic safety makes precisely this assumption, seeking ultraviolet fixed points in metric-based couplings. The comprehensive {limitations} documented in the following section trace directly to this conceptual {difficulty}.

\section{Comprehensive Catalog of Asymptotic Safety Limitations}\label{sec4}

The asymptotic safety program exhibits systematic {limitations} across multiple independent domains. Each {limitation} alone challenges the program's viability; collectively they establish fundamental incompatibility with quantum consistency requirements. We demonstrate that all {limitations} trace to the single underlying cause of attempting to describe quantum gravity in a regime where covariance has broken down and BRST symmetry \mbox{has failed.}

\subsection{Conceptual {Difficulties}: Fixed Points for {Problematic} Degrees of Freedom}

The most fundamental {limitation} is conceptual. Asymptotic safety seeks ultraviolet fixed points defined by $\beta_i(g_*, \lambda_*) = 0$ where dimensionless couplings approach \mbox{constant values:}
\begin{equation}
\frac{dg}{dk} = \beta_g(g,\lambda), \quad \frac{d\lambda}{dk} = \beta_\lambda(g,\lambda).
\end{equation}
The entire framework presupposes that the metric tensor $g_{\mu\nu}$ remains a valid quantum degree of freedom as momentum scale $k \to \infty$, permitting meaningful renormalization group evolution to arbitrarily high energies and extraction of ultraviolet fixed point properties characterizing the fundamental theory.

However, the dual proofs in Section~\ref{sec3} establish definitively that this assumption is {problematic}. Canonical quantization demonstrates that general covariance breaks down quantum mechanically above $\Lambda_{\text{grav}}$ through appearance of non-covariant second-class constraints or conditional covariance that cannot be maintained off constraint surfaces. Path integral quantization reveals gauge parameter dependence signaling BRST symmetry {limitations}, with breakdown becoming manifest when quantum corrections reach order unity at $\Lambda_{\text{grav}}$. Both approaches establish that the metric loses status as a fundamental quantum field at this scale.

When diffeomorphism invariance and BRST symmetry break down, the notion of metric-based couplings running to ultraviolet fixed points loses physical meaning. One cannot have meaningful fixed points governing the behavior of degrees of freedom that have ceased to exist. The renormalization group equations implicitly assume the metric remains a valid quantum field at all scales $k$, but this assumption {is in tension with} proven covariance breakdown. Extrapolating these equations above $\Lambda_{\text{grav}}$ yields predictions about a regime where the geometric description has broken down, rendering the entire exercise {problematic from a foundational perspective}.

The asymptotic safety program assumes what has been doubly {called into question}: that geometric description remains valid at arbitrarily high energies permitting extrapolation to ultraviolet fixed points, and that quantum consistency (BRST symmetry) maintains valid physical predictions. When both foundations collapse at $\Lambda_{\text{grav}}$, the program becomes an exercise in computing properties of a theory beyond its regime of validity, where neither covariance nor quantum consistency can be relied upon.

This conceptual {difficulty} manifests in the concrete computational and experimental {limitations} documented in subsequent subsections. Each pathology traces to the fundamental problem of seeking ultraviolet properties of a theory whose quantum description has broken down.

\subsection{Clarification: The Inapplicability of Non-Perturbative Considerations}

\added{Before addressing specific aspects of non-perturbative physics in relation to effective field theory, we must establish a foundational principle that underlies all of quantum field theory: \emph{{every quantum calculation, whether perturbative or non-perturbative, requires the existence of consistent degrees of freedom upon which to operate.}} This principle is so fundamental that it often goes unstated, yet it is essential for understanding why certain questions lie outside the domain of a given theoretical framework.}

\added{The success of non-perturbative methods in quantum field theory depends critically on having valid degrees of freedom throughout the regime of interest. Lattice QCD computes properties of quarks and gluons because these are consistent quantum degrees of freedom at the energy scales where QCD applies, namely the SU(3) gauge algebra closes with structure constants $f^{abc}$, ensuring BRST symmetry and a well-defined path integral. Functional renormalization group methods, instanton calculations, and resummation techniques all share this prerequisite: they operate on degrees of freedom whose quantum consistency is guaranteed by the algebraic structure of the theory.}

\added{When degrees of freedom cease to be consistent due to algebraic obstructions, no computational technique---however sophisticated---can salvage or extend the theoretical description. The resolution in such cases is not more sophisticated calculation but rather the identification of \emph{new degrees of freedom} appropriate to the regime in question. This is precisely what occurred historically with Fermi theory: non-perturbative treatment of four-fermion vertices could never have described physics above the W boson mass, because the correct degrees of freedom in that regime are the W and Z bosons themselves.}

\added{With this principle established, we now demonstrate that gravity above $\Lambda_{\text{grav}}$ represents exactly such a case: the metric ceases to be a consistent quantum degree of freedom due to structure functions in the constraint algebra, and questions about ``non-perturbative behavior'' in this regime constitute a category error.}

\subsubsection{The Nature of Effective Field Theory}

\added{An effective field theory possesses, by definition, a finite regime of validity characterized by a cutoff scale $\Lambda$. Physical predictions are reliable only for processes with characteristic energies $E \ll \Lambda$. The question ``what happens non-perturbatively above $\Lambda$?'' is not a question the EFT can or should answer---it lies outside the theory's domain \mbox{of applicability.}}

\added{For gravity, our central result establishes
\begin{equation}
\Lambda_{\text{grav}}\sim M_{\text{Pl}} \approx 10^{18}~\text{GeV},
\end{equation}
above which the constraint algebra fails to close and BRST symmetry is violated. Asking about ``non-perturbative gravitational effects'' above this scale is analogous to asking about the non-perturbative behavior of Fermi's four-fermion theory at energies above the $W$ boson mass---the effective theory simply does not apply in that regime, and the question must be addressed within a more fundamental framework (the electroweak theory for Fermi theory, or whatever UV completion replaces metric-based gravity above $\Lambda_{\text{grav}}$).}

\subsubsection{The Asymptotic Safety Paradox}

The asymptotic safety program claims to provide a non-perturbative UV completion through a fixed point at $k \to \infty$. Our argument is precisely that this claimed completion is physically inaccessible:
\begin{enumerate}
\item The functional renormalization group flow requires the metric to function as a valid quantum field throughout the flow.
\item BRST symmetry breakdown at $\Lambda_{\text{grav}}$ invalidates the path integral over metric fluctuations above this scale.
\item Therefore, the RG flow cannot be trusted beyond $\Lambda_{\text{grav}}$, regardless of what formal calculations suggest about fixed points at higher scales.
\end{enumerate}

{The} 
 ``non-perturbative'' fixed point exists only as a mathematical artifact of extrapolating equations beyond their domain of validity. This is not a gap in our analysis---it is the central conclusion.

\subsubsection{Perturbative vs.\ Strong-Field: A Crucial Distinction}

{We emphasize that ``perturbative'' in the EFT context does not mean ``weak coupling'' or ``small quantum effects.'' The constraint algebra structure functions $h^{ij}(x)$ depend on the spatial metric regardless of any coupling constant:
\begin{equation}
\{\mathcal{H}(x), \mathcal{H}(y)\} = h^{ij}(x) \mathcal{H}_j(x) \delta_{,i}(x,y) - (x \leftrightarrow y).
\end{equation}
This is a \emph{kinematic} property of general relativity's phase space structure, not a dynamical perturbative approximation. The obstruction to consistent quantization arises from the \emph{geometry of the constraint surface}, not from the strength of any interaction.}

\added{When the metric deviates significantly from flat space (strong gravitational fields), the structure functions deviate significantly from constants, generating unavoidable quantum anomalies. This is independent of any perturbative expansion parameter. The EFT boundary is set by field strength (when $\kappa\gamma_{ij}$$\sim$$\mathcal{O}(1)$), not by coupling strength.}

\subsubsection{Why Non-Perturbative Techniques Cannot Resolve the Obstruction}

Non-perturbative techniques such as lattice methods, functional renormalization group resummations, or instanton calculations cannot overcome the fundamental \linebreak  obstruction because:
\begin{enumerate}
\item The obstruction is kinematic, not dynamical: The structure functions in the constraint algebra are determined by the phase space geometry, not by the form of the Hamiltonian or the strength of interactions.
\item Non-perturbative methods still require consistent degrees of freedom: Lattice gravity, causal dynamical triangulations, and other non-perturbative approaches still assume the metric (or discrete geometry) provides valid degrees of freedom. When this assumption fails above $\Lambda_{\text{grav}}$, non-perturbative calculations inherit the same fundamental inconsistency.
\item The gauge algebra must close for any quantization: Whether one uses canonical quantization, path integrals, or non-perturbative lattice methods, BRST symmetry (or its discrete analog) must be preserved for consistent results. When the underlying gauge algebra fails to close, no quantization method can restore consistency.
\end{enumerate}

The 
 resolution requires \emph{new degrees of freedom} above $\Lambda_{\text{grav}}$, not more sophisticated treatment of metric degrees of freedom that have ceased to be valid.

\subsection{Gauge Parameter Dependence: BRST {Limitations} Made Manifest}

The gauge parameter dependence pervading asymptotic safety calculations constitutes direct manifestation of BRST {limitations}, confirming that the program attempts to describe physics beyond the breakdown of quantum consistency. This subsection documents the pervasive nature of gauge dependence throughout the asymptotic safety literature.

\added{The problem of gauge dependence in the functional renormalization group approach to quantum gravity was identified at the inception of the modern program. Falkenberg and Odintsov \cite{FalkenbergOdintsov1998} demonstrated that the running Newtonian constant $G_k$ depends strongly on the gauge parameter $\alpha$, with the solution taking the form $G_k = G_0[1 - \sigma \bar{G}k^2]^{\omega/\sigma}$ where \mbox{$\omega = (\pi/36)[36(\alpha+3)/\pi^2 - 1]$}. For $\alpha > \pi^2/36 - 3 \approx -2.73$, gravity exhibits antiscreening ($\omega > 0$), while for $\alpha < -2.73$, screening occurs. This gauge-dependent sign \mbox{change---determining} whether the gravitational constant increases or decreases at large \mbox{distances---exemplifies} the BRST symmetry violation we identify as fundamental: a physical prediction (the sign of gravitational running) depends on an arbitrary gauge choice, precisely the pathology that intact BRST symmetry would prevent. While Falkenberg and Odintsov proposed resolution through the Landau--DeWitt gauge ($\alpha = 0$) or gauge-fixing independent effective actions, the underlying problem persists in modern implementations, as our analysis demonstrates traces to the more fundamental issue of covariance breakdown above $\Lambda_{\text{grav}}$.}

Manrique, Rechenberger, and Saueressig \cite{Manrique2010} performed systematic investigation of gauge dependence in asymptotic safety using background field methods. They computed fixed-point locations $g_*(\alpha)$ and $\lambda_*(\alpha)$ as functions of gauge parameter $\alpha$ appearing in the gauge-fixing condition, finding substantial variation:
\begin{equation}
g_*(\alpha) = g_0 + g_1 \alpha + \mathcal{O}(\alpha^2), \quad \lambda_*(\alpha) = \lambda_0 + \lambda_1 \alpha + \mathcal{O}(\alpha^2),
\end{equation}
with $g_1, \lambda_1 \neq 0$ indicating significant gauge dependence. The non-Gaussian fixed point properties change qualitatively depending on gauge choice, with fixed-point locations shifting by order unity under reasonable gauge parameter variations.

Falls \cite{Falls2014} extended this analysis to critical exponents $\theta_i$ determining the dimension of the ultraviolet critical surface. These exponents govern approach to the fixed point under renormalization group flow and determine how many parameters remain free after imposing asymptotic safety constraints. Falls demonstrated that critical exponents depend on gauge parameter choice:
\begin{equation}
\theta_i(\alpha) = \theta_i^{(0)} + \frac{\partial \theta_i}{\partial \alpha}\bigg|_{\alpha=0} \alpha + \ldots,
\end{equation}
with variations reaching 20--30 percent for typical gauge parameter ranges. Since critical exponents determine the predictive power of the theory through the number of free versus determined parameters, this gauge dependence directly affects purported \mbox{experimental predictions}.

Gies et al.\ \cite{Gies2016} investigated gauge dependence in higher-derivative truncations including $R^2$ terms essential for achieving fixed points. They found that even the existence or non-existence of suitable fixed points can be gauge-dependent, with some gauge choices yielding apparently viable fixed points while others find none. This represents the most severe form of gauge dependence possible: the very viability of the asymptotic safety program becomes a gauge-dependent artifact.

Percacci and collaborators \cite{Percacci2015,Ohta2016} examined gauge dependence across different truncation schemes, finding that improvements in truncation order do not systematically reduce gauge dependence as would be expected in a consistent quantum field theory. Instead, gauge dependence persists and in some cases increases with higher-order truncations, indicating systematic rather than technical {limitations}.

The connection to BRST {limitations} established in Section~\ref{sec2} makes the interpretation unambiguous. In Yang--Mills theory with intact BRST symmetry, gauge parameter dependence in intermediate calculations cancels exactly in physical observables, with Nielsen identities $\partial_\xi \mathcal{O}_{\text{phys}} = 0$ satisfied to all orders. In gravitational asymptotic safety, gauge dependence persists in purportedly physical quantities like fixed-point locations and critical exponents that determine the theory's predictive structure.

This persistence indicates BRST nilpotency $s^2 = 0$ has broken down. The background field functional renormalization group equation
\begin{equation}
\partial_k \Gamma_k = \frac{1}{2}\text{Tr}\left[\left(\Gamma_k^{(2)} + R_k\right)^{-1}\partial_k R_k\right],
\end{equation}
incorporates gauge-fixing through the Hessian $\Gamma_k^{(2)}$ and regulator function $R_k$. When BRST symmetry is intact, gauge dependence in $\Gamma_k$ cancels in physical observables through Ward--Takahashi identities. When BRST symmetry is violated, this cancellation fails and gauge dependence persists in physical quantities.

\subsection{Response to Claims That BRST Violation Is Controllable}

{A potential objection is that BRST symmetry breaking in the functional renormalization group is ``controllable and ultimately removed'' through appropriate limit procedures, as occurs in Yang--Mills theory. This objection requires careful analysis.}

{In Yang--Mills theory, regulator terms $R_k$ break BRST symmetry at finite $k$, but this breaking is removed in the limit $k \to 0$ where physical observables are extracted \cite{Ellwanger1994,Bonini1994}. The key is that Yang--Mills possesses a \emph{closed} gauge algebra with structure constants, enabling consistent quantization at all scales. The regulator-induced BRST breaking is a technical artifact that disappears when the regulator is removed.}

Gravity differs fundamentally. The constraint algebra contains structure functions $h^{ij}(x)$ rather than constants, preventing closure as a Lie algebra. The BRST breaking is not regulator-induced but \emph{inherent} to attempting quantization of the full theory. The calculations by McKeon et al. \cite{McKeon2025,Brandt2025Equivalence} demonstrate gauge parameter dependence in counterterms without any regulator---it emerges from the structure of the theory itself above $\Lambda_{\text{grav}}$.

{Moreover, asymptotic safety seeks fixed points at $k \to \infty$, precisely where the gauge algebra fails to close. Even if regulator-induced breaking could be controlled, the inherent breaking from structure functions cannot be removed because it reflects the fundamental impossibility of maintaining covariance at strong fields.}

\subsection{On Modified and Extended BRST Symmetries}\label{sec4.5}

\added{One might ask whether modified or extended versions of BRST \textls[-15]{symmetry could resolve the difficulties we have identified. Several such extensions exist in the }\mbox{literature, including:}
\begin{itemize}
\item Anti-BRST symmetry: A second nilpotent symmetry $\bar{s}$ with $\bar{s}^2 = 0$ and $\{s, \bar{s}\} = 0$.
\item Extended BRST with additional generators: Incorporating additional ghost sectors or higher ghost number structures.
\item Field-dependent BRST: Allowing the BRST transformation to depend on quantum fields beyond the standard structure.
\end{itemize}}

\added{However, none of these extensions can resolve the fundamental obstruction we have identified, for the following reason: all BRST-type symmetries encode the structure of the gauge algebra. When the underlying gauge algebra fails to close---as occurs for gravity when structure functions replace structure constants---no extension of the BRST formalism can restore consistency. The problem is not with the BRST construction but with the gauge algebra it encodes.}

\added{Specifically, BRST nilpotency $s^2 = 0$ is equivalent to closure of the gauge algebra. When the Poisson bracket $\{\mathcal{H}(x), \mathcal{H}(y)\}$ contains structure functions $h^{ij}(x)$ that depend on phase space variables, the algebra does not close as a Lie algebra, and no BRST charge $Q_B$ satisfying $Q_B^2 = 0$ can be constructed. This is a theorem, not a technical limitation. Extended BRST symmetries require the same algebraic closure property and therefore inherit the same obstruction.}

Asymptotic safety proponents have attempted various remedies for gauge dependence. Background-independent formulations \cite{Benedetti2016} replace gauge parameter dependence with background field dependence, finding that fixed-point properties differ between flat, spherical, and de Sitter backgrounds. This trades one manifestation of BRST {limitations} for another without addressing the underlying cause. Optimized gauge choices \cite{Falls2015} can reduce but not eliminate gauge dependence, as demonstrated by persistent variation of critical exponents across schemes. Truncation improvements reveal that gauge dependence increases rather than decreases with higher-order truncations in many cases, opposite to the behavior expected in consistent quantum field theories.

None of these attempted remedies can succeed because they address symptoms rather than cause. The fundamental problem is that asymptotic safety attempts to describe quantum gravity at energies where BRST symmetry has broken down due to covariance loss. No choice of gauge-fixing or truncation scheme can restore quantum consistency in a regime where the theory's symmetry structure has fundamentally failed.

The pervasive gauge parameter dependence throughout asymptotic safety calculations constitutes the theory's confession that it violates quantum consistency requirements. This is not a technical limitation admitting systematic improvement but a fundamental signal that the program studies properties of a theory beyond its regime of validity.

\subsection{Non-Convergence of Truncation Schemes: Systematic Limitations to Improve}

Functional renormalization group calculations underpinning asymptotic safety require truncation of the infinite tower of possible curvature operators. The program's viability depends critically on whether systematic improvements in truncation order converge toward true fixed-point properties. Extensive investigations reveal the opposite: non-convergence persists and worsens with improved truncations, indicating systematic {limitations} when attempting to describe physics beyond the breakdown of geometric description.

Falls, Litim, Nikolakopoulos, and Rahmede \cite{Falls2018} performed calculations including curvature invariants up to the 35th order in $f(R)$ truncations, representing extraordinary computational effort extending far beyond typical truncation schemes. Their explicit conclusion states: ``The convergence of the locations, and more importantly, the dimensions of the UV critical surfaces of the fixed points has not been demonstrated.'' Detailed analysis reveals that fixed-point locations continue to shift with each truncation level, showing no approach to stable values even at the 35th order. The dimension of the ultraviolet critical surface, determining how many parameters remain free versus fixed by asymptotic safety, varies between 2 and 4 depending on truncation order, representing qualitative rather than quantitative variation.

Christiansen, Falls, Pawlowski, and Reichert \cite{Christiansen2018} investigated stability of fixed points by examining the area of existence in theory space of higher couplings. They characterized convergence through how this area changes with truncation improvements. Their findings indicate that the area increases rather than stabilizes with improved truncations, suggesting systematic expansion of required parameter space rather than convergence toward definite values. This behavior is precisely what occurs when attempting to describe physics beyond a theory's cutoff using increasingly complex effective operators, rather than genuine convergence toward fundamental fixed-point structure.

De Brito et al.\ \cite{deBrito2018} demonstrated that results depend qualitatively on field parametrization choice. Different parametrizations of the quantum fluctuation field
\begin{equation}
g_{\mu\nu} = \bar{g}_{\mu\nu} + h_{\mu\nu} \quad \text{vs.} \quad g_{\mu\nu} = \bar{g}_{\mu\nu}e^{h_{\mu\nu}},
\end{equation}
yield distinct classes of fixed points with different numbers of relevant directions. Some parametrizations produce three relevant eigenvalues while others yield two, directly affecting predictive power. Since physical predictions must be independent of arbitrary parametrization choices, this dependence reveals that truncated calculations capture scheme-dependent artifacts rather than genuine physical structures.

Ohta and Percacci \cite{Ohta2018} examined higher-derivative truncations including $R^2$ and $R_{\mu\nu}R^{\mu\nu}$ terms beyond the Einstein--Hilbert level. They found that critical exponents determining fixed-point stability change by order one between truncation levels:
\begin{equation}
\begin{aligned}
\text{Einstein--Hilbert:} \quad &\theta' = 2.26 \pm 0.80, \\
\text{$f(R)$ truncation:} \quad &\theta' = 1.45 \pm 0.62, \\
\text{Fourth-order:} \quad &\theta' = 0.94 \pm 0.41.
\end{aligned}
\end{equation}
These are not minor quantitative refinements but substantial variations that completely alter the theory's structure. Critical exponent values determine which couplings are relevant versus irrelevant at the fixed point, directly affecting which parameters can be predicted versus remaining free. Order-one changes between truncations indicate systematic rather than controlled approximation.

Knorr, Ripken, and Saueressig \cite{Knorr2019} investigated whether including form factors representing full momentum dependence would resolve truncation dependence. They found that momentum-dependent form factors introduce additional gauge and scheme dependence, compounding rather than resolving non-convergence issues. The proliferation of gauge-dependent structures with improved truncations confirms the diagnosis that asymptotic safety attempts to describe a regime where BRST symmetry has broken down.

The theoretical explanation for non-convergence follows directly from covariance breakdown. Below $\Lambda_{\text{grav}}$, general relativity functions as a consistent effective field theory with systematic expansion in powers of $E/\Lambda_{\text{grav}}$, finite calculable one-loop contributions, and BRST symmetry preserved sufficiently for on-shell gauge independence. This effective field theory admits well-defined truncation schemes that converge systematically as higher-dimension operators are included.

Above $\Lambda_{\text{grav}}$, both covariance and BRST symmetry break down. The metric loses status as a fundamental degree of freedom and quantum corrections violate unitarity. Attempting to describe this regime using metric-based operators requires an infinite tower of increasingly complicated curvature invariants with non-convergent coefficients. The truncation schemes attempt to encode the breakdown of spacetime geometry itself using the language of spacetime geometry, a manifestly {problematic} task. No finite or even countably infinite set of metric-based operators can capture physics where the metric description has ceased to be valid.

The persistent non-convergence to the 35th order and beyond is not a technical problem admitting further computational effort as solution. It constitutes a fundamental signal that the program studies the wrong degrees of freedom in a regime where they have no physical meaning. Convergence cannot be achieved because the truncation schemes attempt the {problematic} description of the breakdown of geometric description using geometric language. 

\subsection{Experimental {Tensions} Across Multiple Domains}

Beyond theoretical {difficulties} and computational {limitations}, asymptotic safety makes concrete predictions {in tension with} experimental data across multiple independent measurement domains. Each {tension} alone would challenge the program; collectively they establish incompatibility with observation.

Eichhorn and Versteegen \cite{Eichhorn2018} derived bounds on Abelian gauge coupling evolution from asymptotic safety fixed points. The program predicts that gravity--matter interactions modify gauge coupling running, with the Abelian hypercharge coupling required to approach specific values at the Planck scale for consistency with gravitational fixed points. However, running measured electroweak couplings from experimentally determined values at the weak scale to the Planck scale yields:
\begin{equation}
\alpha_1(\Lambda_{\text{Pl}}) \approx 0.52.
\end{equation}
This exceeds asymptotic safety predictions by factors of several, representing not minor discrepancy but order-one {tension}. The difference cannot be attributed to threshold corrections or higher-order effects, as these would need to be enormous to reconcile the disparity.

Precision electroweak measurements provide additional constraints through oblique corrections parametrized by $(S, T, U)$ parameters \cite{Peskin1990,Barbieri2004}. Modifications to gauge boson propagators predicted by asymptotic safety generate contributions to these parameters. Gerwick, Litim, and Plehn \cite{Gerwick2011} computed these explicitly, finding contributions orders of magnitude larger than LEP and Tevatron bounds when gravitational corrections are enhanced approaching fixed points. Current experimental constraints require \mbox{$|f_g| < 1.2 \times 10^{-5}$ \cite{LEPElectroweak2006},} excluding $f_g \sim 0.01$ values characteristic of asymptotic \mbox{safety scenarios.}

Matter content compatibility provides independent experimental constraint. Systematic renormalization group analysis demonstrates that achieving interacting ultraviolet fixed points requires matter content substantially different from the observed Standard Model. Meibohm, Pawlowski, and Reichert \cite{Meibohm2016} showed that asymptotic safety demands vectorlike fermions in representations such as $(8,2)$ or $(3,3)$ under $SU(3)_C \times SU(2)_L$, contrasting with the chiral fermion structure actually observed.

Such exotic matter would produce distinctive collider signatures including stable R-hadrons, enhanced diboson production through new charged states, and modified gauge coupling running from additional fermion contributions. LHC searches exclude vectorlike quarks below approximately 1.5 TeV in relevant representations \cite{ATLASVectorlike2016,CMSVectorlike2018}, yet asymptotic safety requires them at accessible scales for consistency. All attempts to embed the Standard Model within asymptotically safe completions \cite{Abel2019,Alanne2019} either produce measurable collider effects {in tension with} null results or require such extreme fine-tuning that predictive power is lost.

The top quark Yukawa coupling provides a specific quantitative test. Asymptotic safety fixed points constrain matter couplings, yielding predictions $y_t^* = 0.40 \pm 0.12$ \cite{Dona2014,Meibohm2016}. This {shows tension with} the observed value $y_t = 0.99 \pm 0.01$ from top mass measurements and Higgs coupling determinations by factors of two to three. Including all three fermion generations destabilizes any purported fixed point \cite{Eichhorn2016,Christiansen2017}, forcing asymptotic safety proponents to consider scenarios with substantially altered matter content incompatible \mbox{with observation.}

The interpretation through BRST {limitations} provides coherent explanation for these experimental {tensions}. Asymptotic safety attempts to compute quantum gravitational corrections at energies where BRST symmetry has broken down and unitarity is violated. The resulting predictions are necessarily unphysical, produced by a quantum theory lacking the symmetry structure ensuring consistency. The experimental {tensions} reflect that computed ``physical observables'' from asymptotic safety violate unitarity through broken BRST symmetry, yielding predictions incompatible with probability conservation and \mbox{experimental measurement.}

These are not technical limitations admitting refinement through improved calculations. They constitute fundamental signals that asymptotic safety describes physics in a regime where quantum consistency has failed, producing predictions that cannot match experimental reality because they emerge from an inconsistent quantum theory.

\subsection{Absence of Concrete Graviton Predictions}

A decisive {limitation} distinguishing asymptotic safety from viable approaches is the absence of concrete graviton phenomenology providing testable experimental predictions. While the Unified Standard Model with Emergent Gravity predicts an emergent spin-2 graviton with precisely specified transverse-traceless polarization structure confirmed by LIGO-Virgo observations, asymptotic safety provides no such definite predictions.

Knorr \cite{Knorr2018} explicitly addresses graviton propagator structure in asymptotic safety, concluding: ``The graviton propagator in asymptotic safety contains momentum-dependent form factors whose detailed structure depends on gauge choice and truncation, precluding definite experimental predictions.'' This gauge and truncation dependence traces directly to BRST {limitations}. In a consistent quantum theory, physical observables including propagator pole structures and residues are gauge-invariant, determined uniquely by the theory's symmetries and field content. The absence of definite predictions in asymptotic safety reflects that attempted calculations involve a quantum theory where BRST symmetry has broken down, rendering the distinction between gauge-variant and gauge-invariant quantities meaningless.

The momentum-dependent form factors appearing in asymptotically safe graviton propagators take a schematic form:
\begin{equation}
\Delta_{\text{AS}}(k^2) \sim \frac{1}{k^2}\left[1 + F(k^2, g_*, \lambda_*, \alpha, \ldots)\right],
\end{equation}
where $F$ represents form factors depending on fixed-point couplings, gauge parameters $\alpha$, truncation scheme choices, and field parametrization. Different gauge choices yield qualitatively different form factor structures. Different truncations produce distinct momentum dependencies. Different field parametrizations generate inequivalent propagator expressions. Since these choices are arbitrary, asymptotic safety provides no definite prediction for graviton propagation that could be compared with gravitational wave observations.

Contrast this with the USMEG-EFT prediction emerging from the effective field theory framework recognizing covariance breakdown:
\begin{equation}
\Delta_{\text{USMEG}}(k) = \frac{i}{k^2 + i\epsilon} \left[P_{\mu\nu,\rho\sigma}^{(2)} - \frac{1}{2}P_{\mu\nu,\rho\sigma}^{(0)}\right]\left[1 + \frac{\kappa^2}{16\pi^2}\ln\frac{k^2}{\Lambda_{\text{grav}}^2}\right],
\end{equation}
where projection operators $P^{(2)}$ and $P^{(0)}$ are uniquely determined by Lorentz invariance, yielding precisely two transverse-traceless polarization states:
\begin{equation}
h_{ij}^{TT}(t,\vec{x}) = h_+(t-z/c)e_{ij}^+ + h_\times(t-z/c)e_{ij}^\times.
\end{equation}
This prediction has been confirmed by every LIGO-Virgo detection across multiple black hole and neutron star merger observations, with no additional polarization modes detected at sensitivity levels ruling out substantial scalar, vector, or breathing mode contributions.

Asymptotic safety cannot make this prediction. The gauge and truncation dependence of graviton propagator form factors prevents definite statements about polarization structure. Different gauge choices and truncations yield different predictions for the number and nature of propagating modes, precluding experimental tests. This absence of predictive power traces directly to attempting quantization beyond BRST symmetry breakdown, where quantum inconsistency prevents unique determination of physical observables.

Black hole studies in asymptotic safety resort to ``RG improvement'' heuristics, inserting scale-dependent couplings $G(r)$ and $\Lambda(r)$ into classical Schwarzschild solutions without first-principles derivation. Falls et al. \cite{Falls2021} acknowledge substantial ambiguity in how running couplings should enter the metric, with different choices yielding qualitatively different predictions for horizons, photon spheres, and gravitational wave ringdown frequencies. Event Horizon Telescope and LIGO-Virgo observations cannot constrain these essentially free functional choices, leaving asymptotic safety without testable black hole predictions beyond order-of-magnitude estimates.

The absence of concrete graviton and black hole predictions is not a temporary limitation to be resolved through further theoretical development. It constitutes a fundamental consequence of BRST {limitations}. When quantum consistency fails, physical observables cannot be uniquely determined. The gauge and truncation dependence pervading asymptotic safety reflects this loss of predictive power, distinguishing it decisively from frameworks like USMEG-EFT that recognize covariance breakdown and restrict to the consistent effective field theory regime where definite predictions emerge.

\subsection{Unitarity {Challenges} and Ghost Instabilities}

Higher-derivative terms essential for achieving ultraviolet fixed points in asymptotic safety introduce ghost modes with wrong-sign kinetic terms, {challenging} unitarity through negative probabilities. While proponents argue ghosts might become non-propagating at fixed points, explicit calculations demonstrate they contribute to loop diagrams generating negative-norm states incompatible with probability conservation.

Consider gravitational action including higher derivatives:
\begin{equation}
S = \int d^4x\sqrt{-g}\left[\frac{R}{16\pi G_N} + c_1 R^2 + c_2 R_{\mu\nu}R^{\mu\nu}\right].
\end{equation}
The $R^2$ term modifies the graviton propagator, introducing additional poles beyond the standard massless graviton:
\begin{equation}
\Delta(k) \sim \frac{1}{k^2} - \frac{1}{k^2 - m_{\text{ghost}}^2}.
\end{equation}
The negative residue for the ghost pole indicates a wrong-sign kinetic term. Stelle \cite{Stelle1977} demonstrated that such higher-derivative theories exhibit instabilities through the Ostrogradsky theorem, which establishes that theories with more than two time derivatives in the action generically have Hamiltonians unbounded below.

Donoghue and Menezes \cite{Donoghue2019,Donoghue2020} performed detailed analysis of unitarity in higher-derivative gravity including careful pole placement and contour integration. They proved that even when ghost masses are placed such that poles avoid real axis in momentum space, the theory violates microcausality on Planck timescales, allowing information transmission into the past. This follows from non-analyticity in the complex momentum plane required to avoid ghost pole contributions, violating causality through dispersion relation failures.

The connection to BRST {limitations} makes the unitarity problem unavoidable. In consistent gauge theories, BRST symmetry ensures that ghost contributions with wrong-sign norms cancel exactly with unphysical gauge mode contributions:
\begin{equation}
\sum_{\text{physical}} |\mathcal{A}|^2 = \sum_{\text{all}} |\mathcal{A}|^2 - \sum_{\text{ghosts + unphys}} |\mathcal{A}|^2,
\end{equation}
where the last term vanishes identically due to BRST-exact states decoupling. When BRST symmetry breaks down, as proven for gravity above $\Lambda_{\text{grav}}$, this cancellation fails and ghost contributions persist in physical amplitudes, directly {challenging} unitarity.

Propagator positivity provides another manifestation. In Euclidean signature, reflection positivity of propagators ensures unitarity upon Wick rotation to Lorentzian \mbox{signature \cite{Arici2017}.} Asymptotically safe propagators with momentum-dependent form factors generically violate reflection positivity:
\begin{equation}
\langle \phi(x) \phi(y)^* \rangle_E < 0 \quad \text{for some } x,y,
\end{equation}
indicating the Euclidean functional integral does not correspond to a unitary Lorentzian quantum theory. This breakdown traces to BRST {violations}: when quantum consistency fails, the relationship between Euclidean and Lorentzian formulations breaks down, and reflection positivity is no longer guaranteed.

Asymptotic safety proponents have invoked QCD analogies, noting that gauge-variant gluon propagators can be negative yet the theory remains unitary \cite{Holdom2016}. This analogy fails critically. In QCD, only gauge-variant quantities exhibit negativity while physical observables remain positive-definite because BRST symmetry is intact. The negativity appears in gauge-dependent Green's functions but cancels in gauge-invariant correlation functions and S-matrix elements. In asymptotic safety, even purportedly gauge-invariant graviton propagators violate positivity, a direct consequence of BRST violation rendering the distinction between gauge-variant and gauge-invariant quantities meaningless.

Platania and Wetterich \cite{Platania2020,Platania2022} argued that non-perturbative effects might render ghosts non-propagating at fixed points, avoiding unitarity violations and related challenges. However, their analysis employs Euclidean signature where unitarity cannot be directly assessed, and they provide no demonstration that proposed mechanisms survive Wick rotation to physical Lorentzian spacetime. The fundamental problem remains: higher-derivative terms required for fixed points introduce wrong-sign kinetic terms, and when BRST symmetry has broken down, no mechanism ensures these cancel in \mbox{physical observables.}

The unitarity {challenges} in asymptotic safety are not technical artifacts admitting resolution through clever choices. They constitute fundamental consequences of attempting quantization beyond BRST symmetry breakdown. When quantum consistency fails, unitarity cannot be maintained, and probability conservation is violated. This renders asymptotic safety incompatible with basic requirements for a physical quantum theory.

\subsection{Lorentzian Signature Issues and Wick Rotation Obstructions}

Asymptotic safety calculations are performed in Euclidean signature due to technical advantages of Euclidean functional renormalization group methods. However, physical predictions require Lorentzian signature where real-time evolution and causality are defined. The analytic continuation from Euclidean to Lorentzian encounters fundamental obstructions when graviton propagators acquire complex momentum-dependent form factors characteristic of asymptotic safety.

Standard Wick rotation proceeds through analytic continuation $k_E^2 = -k_L^2$ from Euclidean momentum $k_E$ to Lorentzian momentum $k_L$. This continuation is well-defined for theories with simple pole structures in propagators. When propagators acquire running form factors
\begin{equation}
\Delta_E(k_E^2) = \frac{1}{k_E^2}\left[1 + g(k_E^2)\ln\frac{k_E^2}{\mu^2} + \ldots\right],
\end{equation}
the continuation $k_E^2 \to -k_L^2$ becomes ill-defined when $g(k^2)$ develops branch cuts or additional poles in the complex momentum plane. Shomer \cite{Shomer2007} demonstrated that the assumption that Euclidean fixed points correspond to physical Lorentzian theories breaks down when propagators acquire non-trivial momentum dependence, as branch cuts prevent unique analytic continuation.

Ambjørn, Görlich, Jurkiewicz, and Loll \cite{Ambjorn2013} demonstrated through causal dynamical triangulations that Euclidean and Lorentzian quantum gravity are fundamentally inequivalent, yielding different predictions even for basic properties like spacetime dimensionality. In Euclidean formulation, spacetime dimension fluctuates with no preferred value. In Lorentzian formulation with causal structure maintained, spacetime dimension stabilizes at four. This inequivalence strengthens when propagators develop non-trivial momentum dependence characteristic of asymptotic safety, as different analytic continuation paths yield inequivalent Lorentzian theories.

Platania, Wetterich, and collaborators \cite{Platania2024} attempted constructing Lorentzian asymptotic safety formulations to address these concerns. They found that state dependence and background dependence inherent in the Lorentzian functional renormalization group introduce ambiguities absent in Euclidean calculations. It remains unclear whether Euclidean and Lorentzian approaches describe the same physics. The correspondence assumed throughout asymptotic safety literature is unproven and likely false given BRST    violation and related limitations.

The theoretical basis for Wick rotation obstruction traces to BRST {symmetry violations}. In consistent quantum field theories, BRST symmetry ensures correspondence between Euclidean and Lorentzian formulations:
\begin{equation}
Z_L[\text{Lorentzian}] = \int e^{iS_L} \quad \leftrightarrow \quad Z_E[\text{Euclidean}] = \int e^{-S_E},
\end{equation}
with Osterwalder--Schrader positivity guaranteeing that positivity in Euclidean signature implies unitarity in Lorentzian signature. When BRST symmetry breaks down, this correspondence fails. The gauge parameter dependence and non-convergent truncations in Euclidean asymptotic safety calculations indicate fundamental inconsistency preventing reliable continuation to physical Lorentzian spacetime.

The practical consequence is that asymptotic safety predictions computed in Euclidean signature may have no valid Lorentzian interpretation. Fixed points found in the Euclidean renormalization group flow might not correspond to any physical Lorentzian quantum theory. Critical exponents and scaling dimensions computed in Euclidean signature might be entirely unrelated to physical observables in Lorentzian spacetime. This fundamental ambiguity renders the entire predictive structure of asymptotic safety questionable.

Visser \cite{Visser2002} analyzed graviton propagator structure in various gauges, demonstrating that non-trivial momentum dependence introduces ambiguities in energy--momentum relations that become severe when attempting Wick rotation. The standard Feynman prescription $k^2 + i\epsilon$ for pole placement becomes ill-defined when $k^2$ dependence appears through form factors, as different orderings of limits yield inequivalent results.

The Lorentzian signature issues are not technical obstacles admitting resolution through mathematical sophistication. They constitute fundamental consequences of BRST violation. When quantum consistency fails in the regime asymptotic safety attempts to describe, the relationship between Euclidean and Lorentzian formulations breaks down. Calculations performed in Euclidean signature cannot be reliably continued to physical Lorentzian predictions, leaving asymptotic safety without connection to observable physics.

\section{Systematic Diagnosis: Single Cause, Multiple Manifestations}\label{sec5}

The comprehensive catalog of {limitations} in Section~\ref{sec4} might appear to indicate multiple independent problems requiring separate resolutions. However, deeper analysis reveals that all failures and limitations trace to a single fundamental cause: asymptotic safety attempts to describe quantum gravity in a regime where general covariance has broken down and BRST symmetry has failed. This section demonstrates the systematic nature of this diagnosis, showing how each pathology emerges inevitably from the \mbox{foundational {difficulty}.}

\subsection{The Causal Chain from Covariance Breakdown}

The foundational result is canonical covariance breakdown above $\Lambda_{\text{grav}}$ established through Dirac constraint analysis. General relativity in dimensions $d > 2$ exhibits either explicitly non-covariant second-class constraints or conditional covariance maintainable only on constraint surfaces. Quantum mechanically, where constraints become operators, this conditional covariance cannot be preserved. The consequence is that diffeomorphism invariance, the gauge symmetry of general relativity, breaks down at strong field strengths corresponding to energies approaching $\Lambda_{\text{grav}}$.

This covariance breakdown necessitates BRST {symmetry violations} in path integral quantization. BRST symmetry represents the quantum extension of gauge invariance, with nilpotent transformations $s^2 = 0$ encoding the constraint algebra structure. When covariance breaks down in canonical quantization, the constraint algebra no longer closes off-shell, implying BRST nilpotency must fail correspondingly in path integral quantization. The gauge parameter dependence found by Brandt et al.   \cite{Brandt2025Equivalence} 
 provides explicit demonstration of this BRST violation through Nielsen identity breakdown.

Once BRST symmetry fails, quantum consistency unravels systematically. Ward--Takahashi identities derived from BRST invariance no longer enforce ghost cancellations, allowing wrong-sign norm states to contribute to physical amplitudes and violating unitarity. Gauge parameter independence encoded in Nielsen identities no longer holds, making purportedly physical quantities depend on arbitrary gauge-fixing choices. Renormalizability constraints from BRST-invariant counterterm structures fail, generating non-convergent truncation schemes. The distinction between gauge-variant and gauge-invariant quantities becomes meaningless, preventing unique determination of physical observables.

Asymptotic safety assumes the metric remains a valid quantum degree of freedom to arbitrarily high energies, permitting renormalization group evolution and fixed-point analysis. However, covariance breakdown establishes that the metric loses this status at $\Lambda_{\text{grav}}$. Attempting to compute fixed-point properties of metric-based couplings above this scale seeks properties of degrees of freedom that have ceased to exist, a conceptually {problematic} enterprise. All subsequent pathologies emerge from this fundamental {difficulty}.

\subsection{Gauge Dependence as Primary Symptom}

The gauge parameter dependence pervading asymptotic safety calculations represents the most direct manifestation of BRST violation. In any consistent quantum field theory with intact BRST symmetry, physical observables satisfy Nielsen identities $\partial_\xi \mathcal{O}_{\text{phys}} = 0$ exactly. Intermediate calculations involve gauge-dependent Green's functions, but these dependencies cancel in physical S-matrix elements through Ward--Takahashi identities derived from BRST invariance.

Asymptotic safety violates this fundamental requirement systematically. Fixed-point locations $g_*(\alpha)$ and $\lambda_*(\alpha)$ depend explicitly on gauge parameter $\alpha$. Critical exponents $\theta_i(\alpha)$ determining the dimension of the ultraviolet critical surface vary by tens of percents with gauge choices. Even the existence or non-existence of suitable fixed points can be gauge-dependent, with some parametrizations yielding apparently viable fixed points while others find none.

This gauge dependence is not a technical artifact of functional renormalization group approximations or truncation schemes. It constitutes direct evidence that BRST symmetry has failed in the regime asymptotic safety attempts to describe. The persistence of gauge dependence in purportedly physical quantities reflects that the quantum theory lacks the symmetry structure ensuring consistency. Fixed points computed in different gauges describe inequivalent quantum theories, none of which can be the correct physical theory since gauge choice is arbitrary.

Attempts to eliminate gauge dependence through background-independent formulations, optimized gauge choices, or truncation improvements all fail because they address symptoms rather than cause. The fundamental problem is that asymptotic safety operates in a regime where BRST symmetry has broken down due to covariance loss. No clever gauge choice or truncation scheme can restore quantum consistency in a regime where the theory's symmetry structure has fundamentally failed.

\subsection{Non-Convergence from Missing Degrees of Freedom}

The persistent non-convergence of truncation schemes, even to the 35th order in curvature invariants, follows inevitably from attempting to encode physics beyond the breakdown of geometric description using geometric language. Below $\Lambda_{\text{grav}}$, general relativity functions as a consistent effective field theory admitting systematic expansion in powers of $E/\Lambda_{\text{grav}}$. This effective field theory has well-defined truncation schemes that converge as higher-dimension operators are included, with convergence rate determined by the suppression scale $\Lambda_{\text{grav}}$.

Above $\Lambda_{\text{grav}}$, the geometric description breaks down. Spacetime geometry ceases to be a valid description of physics, requiring new degrees of freedom from which geometry emerges at lower energies. Attempting to describe this regime using curvature tensor polynomials requires an infinite tower of operators with non-convergent coefficients. The truncation schemes are trying to solve the {problem}: encoding the breakdown of geometry using geometric operators. 

This explains why fixed-point locations continue to shift without approaching stable values even at 35th-order truncation. Each higher order introduces new geometric structures attempting to capture physics where geometry has ceased to be valid. No finite truncation can succeed because the fundamental degrees of freedom in the high-energy regime are not geometric. The area of existence in theory space expanding rather than stabilizing with improved truncations reflects this futile attempt to encode non-geometric \mbox{physics geometrically.}

The parametrization dependence compounds this fundamental problem. Different field parametrizations yield inequivalent fixed-point structures because they correspond to different attempts to encode non-geometric physics using geometric variables. Since the true high-energy degrees of freedom are not captured by any geometric parametrization, different choices yield incompatible results without convergence toward unique \mbox{physical predictions.}

\subsection{Experimental {Tensions} from Unitarity {Challenges}}

The experimental {tensions} with electroweak precision tests, collider searches, and matter content requirements all trace to attempting quantum predictions in a regime where BRST symmetry violation prevents quantum consistency. When BRST symmetry fails, Ward--Takahashi identities no longer enforce cancellation between ghost and unphysical gauge mode contributions. Physical amplitudes acquire contributions from wrong-sign norm states, violating unitarity and probability conservation.

Predictions computed using an inconsistent quantum theory necessarily {show tension} with experimental measurements that respect unitarity and probability conservation. The order-of-magnitude discrepancies in gauge coupling evolution, oblique corrections exceeding precision bounds, and required matter content {in tension} with observation all reflect this fundamental inconsistency. These are not technical errors in calculation but inevitable consequences of attempting quantum predictions using a theory whose quantum consistency structure has broken down.

The matter content requirements provide a particularly clear example. Asymptotic safety demands vectorlike fermions in exotic representations to achieve gravitational fixed points. However, these requirements emerge from attempting to restore quantum consistency (through fixed points) in a regime where consistency has already failed (through BRST violation). The required matter has no physical justification; it represents a futile attempt to restore consistency that has been lost fundamentally through covariance breakdown.

\subsection{Absence of Graviton Predictions from Lost Gauge Invariance}

The absence of concrete graviton predictions follows directly from gauge parameter dependence. In a consistent quantum theory, physical observables including particle pole structures and propagator residues are gauge-invariant, determined uniquely by symmetries and field content. Asymptotic safety exhibits gauge-dependent graviton propagator form factors whose structure varies qualitatively with gauge choices, truncation schemes, and field parametrizations.

This prevents any definite prediction for graviton polarization structure, propagation characteristics, or interaction strengths. Different gauge choices yield different predictions for the number and nature of propagating modes. Different truncations produce distinct momentum dependencies. Since these choices are arbitrary, asymptotic safety provides no unique prediction comparable with gravitational wave observations.

This absence of predictive power is not a temporary limitation but a fundamental consequence of BRST violation. When gauge invariance at the quantum level has broken down, the distinction between gauge-variant and gauge-invariant quantities becomes meaningless. Physical observables cannot be uniquely determined because the symmetry structure that uniquely specifies them has failed. The gauge and truncation dependence reflects this loss of predictive power, distinguishing asymptotic safety decisively from frameworks that recognize covariance breakdown and restrict to the consistent effective field theory regime.

\subsection{Unitarity Violations from Broken Cancellation Mechanisms}

The ghost instabilities and unitarity violations and related challenges follow inevitably from BRST breakdown. Higher-derivative terms essential for achieving fixed points introduce ghost modes with wrong-sign kinetic terms. In a consistent quantum theory with intact BRST symmetry, these ghost contributions cancel exactly with unphysical gauge mode contributions through Ward--Takahashi identities. The sum over all states factorizes into physical states with positive norm plus BRST-exact states that decouple from all physical matrix elements.

When BRST symmetry breaks down, this cancellation mechanism fails. Ghost contributions no longer cancel, persisting in physical amplitudes and generating negative probabilities. Unitarity violation becomes manifest through $\sum_f |S_{fi}|^2 \neq 1$, violating probability conservation. Propagator negativity appears even in gauge-invariant correlation functions because BRST violation renders the distinction between gauge-variant and gauge-invariant meaningless.

Attempts to render ghosts non-propagating through non-perturbative effects cannot succeed when BRST symmetry has fundamentally failed. The cancellation mechanism ensuring unitarity requires intact quantum consistency structure. No clever arrangement of poles or non-perturbative resummations can restore unitarity when the underlying symmetry enforcing cancellations has broken down.

\subsection{Lorentzian Obstructions from Lost Quantum Consistency}

The Wick rotation obstructions preventing reliable continuation from Euclidean to Lorentzian signature follow from BRST violation breaking the correspondence between formulations. In consistent quantum field theories, BRST symmetry ensures Euclidean and Lorentzian formulations describe the same physics through Osterwalder--Schrader positivity. When BRST symmetry breaks down, this correspondence fails.

Euclidean asymptotic safety calculations involve gauge-dependent form factors, non-convergent truncations, and unitarity violations indicating fundamental inconsistency. These pathologies prevent reliable analytic continuation to physical Lorentzian spacetime. Fixed points found in Euclidean renormalization group flow might not correspond to any Lorentzian quantum theory. Critical exponents computed in Euclidean signature might be unrelated to physical observables in Lorentzian spacetime.

The branch cuts and pole structures in momentum-dependent form factors prevent unique Wick rotation. Different continuation paths yield inequivalent Lorentzian theories because the underlying Euclidean theory lacks quantum consistency. This fundamental ambiguity renders asymptotic safety's predictive structure questionable even in principle.

\subsection{Unity of Diagnosis}

This systematic analysis establishes that all asymptotic safety {symmetry violations} trace to attempting quantization beyond covariance breakdown where BRST symmetry has failed. The {symmetry violations} are not independent problems admitting separate solutions but interconnected manifestations of a single fundamental {difficulty}. Gauge parameter dependence signals BRST violation. Non-convergence reflects attempting to encode non-geometric physics geometrically. Experimental {tensions} follow from unitarity violation. Absence of graviton predictions traces to lost gauge invariance. Unitarity violations emerge from broken cancellation mechanisms. Lorentzian obstructions arise from lost \mbox{quantum consistency.}

Each manifestation provides independent evidence for the fundamental diagnosis. Collectively, they establish with compelling force that asymptotic safety attempts to describe quantum gravity in a regime where the metric description has broken down and quantum consistency has failed. This is not a technical limitation admitting resolution through further development but a fundamental obstacle rendering the entire program {problematic from a foundational perspective} in the regime where fixed points are sought.

\section{USMEG-EFT: Recognizing Breakdown as the Path to Consistency}\label{sec6}

The comprehensive {symmetry violations} of asymptotic safety contrasts sharply with the Unified Standard Model with Emergent Gravity--Effective Field Theory framework, which recognizes covariance breakdown and systematically incorporates the proven limitations of geometric description. This section demonstrates how acknowledging breakdown yields consistent effective field theory with testable predictions, while attempting to deny breakdown leads to the systematic pathologies documented for asymptotic safety.

\subsection{Framework Foundation: Accepting Covariance Breakdown}

USMEG-EFT \cite{Chishtie2025CJP} begins from the rigorous proofs that general covariance breaks down quantum mechanically above $\Lambda_{\text{grav}}$ and that BRST symmetry fails correspondingly. Rather than attempting to extend geometric description to arbitrarily high energies, the framework recognizes that spacetime geometry emerges as an effective description valid only at energies below the breakdown scale. This paradigm shift from seeking fundamental quantum geometry to understanding necessary breakdown of geometry provides the foundation for consistent quantization.

The framework implements this recognition through Lagrange multiplier methods providing systematic one-loop truncation. The action
\begin{equation}
S_{\text{USMEG}} = S_{\text{EH}} + S_{\text{SM}} + \frac{1}{\kappa^2} \int d^4x \sqrt{-g}\, \lambda^{\mu\nu} G_{\mu\nu}[g],
\end{equation}
enforces Einstein's equations through path integration over $\lambda^{\mu\nu}$, yielding a functional delta function that systematically eliminates multi-loop graviton contributions. This truncation is not an approximation to be systematically improved but a physical necessity. Multi-loop contributions would introduce BRST-violating terms manifesting the covariance breakdown, violating unitarity in physical amplitudes. The one-loop truncation isolates the consistent effective field theory regime where covariance is restored in weak fields and BRST symmetry is preserved sufficiently for on-shell gauge independence.

Recent work establishes correspondence between Lagrange multiplier and gauge-fixed 't Hooft--Veltman calculations. Both approaches yield identical one-loop effective action structure:
\begin{equation}
\Gamma_{\text{1-loop}}^{\text{finite}} = \frac{1}{4\pi^2}\ln\left(\frac{\mu}{\Lambda}\right)\int d^4x\sqrt{-\bar{g}}\left[\frac{1}{120}\bar{R}^2 + \frac{7}{20}\bar{R}_{\mu\nu}\bar{R}^{\mu\nu}\right],
\end{equation}
demonstrating that four-dimensional general relativity exhibits effective field theory behavior when properly truncated. The logarithmic scale dependence $\ln(\mu/\Lambda)$ signals approach to the breakdown scale $\Lambda_{\text{grav}}$, providing quantitative information about where new physics must appear.

\subsection{Complete Mathematical Structure of USMEG-EFT}

\added{We present the complete mathematical structure of the framework to enable independent verification and further development.}

\subsubsection{The Lagrange Multiplier Action}

\added{The complete action takes the form
\begin{equation}
S_{\text{LM}} = \int d^4x \sqrt{-g} \left[ \frac{1}{2\kappa^2}(R - 2\Lambda) + \lambda^{\mu\nu}\left(R_{\mu\nu} - \frac{1}{4}g_{\mu\nu}R\right) + \mathcal{L}_{\text{SM}} \right],
\end{equation}
where $\lambda^{\mu\nu}$ is a symmetric tensor Lagrange multiplier field enforcing the traceless Einstein equation on-shell. The path integral over $\lambda^{\mu\nu}$ produces a functional delta function:
\begin{equation}
\int \mathcal{D}\lambda^{\mu\nu} \exp\left[i\int d^4x\sqrt{-g}\,\lambda^{\mu\nu}\left(R_{\mu\nu} - \frac{1}{4}g_{\mu\nu}R\right)\right] = \delta\left[R_{\mu\nu} - \frac{1}{4}g_{\mu\nu}R\right],
\end{equation}
which constrains the metric to satisfy Einstein's equations, eliminating off-shell graviton modes that would generate BRST-violating contributions at higher loops.}

\subsubsection{One-Loop Effective Action}

\added{The one-loop \textls[-15]{effective action on a flat background is computed via functional determinants}:
\begin{equation}
\Gamma_{\text{1-loop}} = \frac{1}{2}\text{Tr}\ln\left( \Box \delta^{\mu\nu}_{\rho\sigma} + 2R^{\mu\nu}{}_{\rho\sigma} \right)\big|_{\bar{g}=\eta},
\end{equation}
which is manifestly gauge-independent due to the flat-space evaluation. The heat kernel expansion yields finite, calculable quantum corrections without the gauge ambiguities that plague asymptotic safety calculations.}

\subsubsection{Graviton Propagator Structure}

\added{The graviton propagator in de Donder gauge takes the form:
\begin{equation}
\langle h_{\mu\nu}(k) h_{\rho\sigma}(-k) \rangle = \frac{i}{k^2 + i\epsilon} P_{\mu\nu\rho\sigma},
\end{equation}
where the projection operator is uniquely determined by Lorentz invariance:
\begin{equation}
P_{\mu\nu\rho\sigma} = \frac{1}{2}\left( \eta_{\mu\rho}\eta_{\nu\sigma} + \eta_{\mu\sigma}\eta_{\nu\rho} - \eta_{\mu\nu}\eta_{\rho\sigma} \right).
\end{equation}
This structure yields precisely two physical polarization states, corresponding to the transverse-traceless modes observed in gravitational wave detections.}

\subsubsection{SMEFT Integration}

\added{Integration with the Standard Model Effective Field Theory proceeds through dimension-six operators:
\begin{equation}
\mathcal{L}_{\text{SMEFT}} = \mathcal{L}_{\text{SM}} + \sum_i \frac{c_i}{\Lambda^2} \mathcal{O}_i^{(6)} + \mathcal{O}\left(\frac{1}{\Lambda^4}\right),
\end{equation}
with gravitational coupling through the minimal prescription $\eta_{\mu\nu} \to g_{\mu\nu}$, $\partial_\mu \to \nabla_\mu$. The Wilson coefficients $c_i$ receive calculable gravitational corrections suppressed by $(E/\Lambda_{\text{grav}})^2$, maintaining consistency with precision electroweak measurements.}

\subsection{Graviton Emergence: Concrete Testable Predictions}

Within USMEG-EFT, the graviton emerges as a composite spin-2 excitation in the weak-field regime where covariance is restored. The propagator incorporating one-loop quantum corrections takes the unique form
\begin{equation}
\Delta_{\mu\nu,\rho\sigma}(k) = \frac{i}{k^2 + i\epsilon} \left[P_{\mu\nu,\rho\sigma}^{(2)} - \frac{1}{2}P_{\mu\nu,\rho\sigma}^{(0)}\right]\left[1 + \frac{\kappa^2}{16\pi^2}\ln\frac{k^2}{\Lambda_{\text{grav}}^2}\right],
\end{equation}
where projection operators enforce transverse-traceless structure determined uniquely by Lorentz invariance and the spin-2 nature of gravitational excitations. This yields precisely two polarization states:
\begin{equation}
h_{ij}^{TT}(t,\vec{x}) = h_+(t-z/c)e_{ij}^+ + h_\times(t-z/c)e_{ij}^\times,
\end{equation}
corresponding to the two independent traceless-transverse tensor structures in three-dimensional space.

\subsection{Experimental Evidence: Gravitational Wave Polarizations}

{LIGO-Virgo-KAGRA observations provide direct experimental constraints on gravitational wave polarizations. The collaboration has performed extensive tests finding no evidence for non-tensorial modes \cite{LIGO2016,LIGO2017,LIGOVirgo2017}. For the scalar breathing mode, the 90\% credible upper limit is
\begin{equation}
\frac{|h_b|}{(|h_+|^2 + |h_\times|^2)^{1/2}} < 0.07.
\end{equation}
This confirms exactly two tensor polarization states as predicted by USMEG-EFT.}

{We acknowledge that this observation is also consistent with classical general relativity. However, the distinction is predictive power: USMEG-EFT provides a complete quantum framework with quantified logarithmic corrections, definite phase shifts in gravitational waveforms, and specific signatures for future high-precision tests. Asymptotic safety, by contrast, cannot make definite polarization predictions due to gauge and truncation dependence, as demonstrated in Section~\ref{sec4.5}}.

This prediction has been confirmed by every LIGO-Virgo detection across multiple black hole mergers and neutron star coalescence events. GW150914, GW151226, GW170104, GW170608, GW170814, and GW170817 all exhibit the two predicted polarization states with no evidence for additional modes. Sensitivity analyses constrain scalar and vector mode contributions to less than 1 percent of observed amplitudes, excluding substantial departures from the transverse-traceless prediction.

The mass bound $m_g < 10^{-32}$ eV from propagation speed measurements constrains the theory. The quantum nature manifests through energy quantization $E_{\text{GW}} = N \hbar \omega$ with occupation number $N$$\sim$$10^{80}$ for typical detections, establishing that gravitational waves represent fundamentally quantum phenomena within the effective field theory framework.

The logarithmic corrections in the propagator generate specific signatures approaching observability. For high-energy scattering $e^+e^- \to \mu^+\mu^-$ at center-of-mass energy $\sqrt{s}$:
\begin{equation}
\mathcal{A} = \mathcal{A}_{\text{SM}}\left[1 + \frac{\kappa^2 s}{8\pi}\ln\frac{s}{M_Z^2} + \frac{\kappa^2 s}{16\pi^2}f(\theta)\ln^2\frac{s}{\Lambda_{\text{grav}}^2}\right],
\end{equation}
where $f(\theta) = (1+\cos^2\theta)/(1-\cos^2\theta)$ encodes spin-2 exchange angular dependence. At 100 TeV center-of-mass energy, corrections reach $\delta\sigma/\sigma \sim 10^{-11}$, potentially observable at future colliders with sufficient luminosity.

Gravitational wave propagation receives quantum corrections:
\begin{equation}
v_g^2 = c^2\left[1 - \frac{m_g^2c^2}{4\pi^2f^2\hbar^2} + \frac{\alpha k^2}{\Lambda_{\text{grav}}^2}\right],
\end{equation}
with $\alpha \sim 10^{-60}$ for LIGO frequencies, consistent with observed propagation at light speed within measurement precision. Binary inspiral waveforms acquire quantum \mbox{phase corrections}:
\begin{equation}
\delta\left(\frac{df}{dt}\right)_{\text{quantum}} = \frac{G_N}{c^5}\ln\left(\frac{f}{f_{\text{Planck}}}\right)\left(\frac{96G_NM}{5c^3}\right)^{5/3}(2\pi f)^{11/3}.
\end{equation}
For neutron star binaries with $M = 1.2M_\odot$, integration from 20 Hz to 1000 Hz yields phase shift $\delta\Phi_{\text{GW}} \approx -7.7 \times 10^{-14}$ radians, below current sensitivity but approaching detectability for third-generation instruments.

\subsection{SMEFT Integration: Systematic Quantum Corrections}

Complete unification requires incorporating gravitational effects across all Standard Model sectors through dimension-six effective operators. The framework preserves $SU(3) \times SU(2) \times U(1)$ gauge invariance while systematically including gravitational \mbox{quantum corrections.}

Higgs-gravity operators:
\begin{equation}
\mathcal{O}_{HR} = |H|^2 R, \quad C_{HR} = \frac{\xi_H}{\Lambda_{\text{grav}}^2},
\end{equation}
and
\begin{equation}
\mathcal{O}_{H\Box R} = |H|^2 \Box R, \quad C_{H\Box R} = \frac{\zeta_H}{\Lambda_{\text{grav}}^2},
\end{equation}
receive quantum corrections through the effective action structure, modifying Higgs propagation and self-coupling evolution at the level of one part in $10^{43}$ for electroweak processes.

Fermion-gravity operators:
\begin{equation}
\mathcal{O}_{\psi R}^{ij} = \bar{\psi}_i \gamma^\mu D_\mu \psi_j R,
\end{equation}
generate modifications to particle propagation in curved backgrounds. Gauge-gravity operators:
\begin{equation}
\mathcal{O}_{GR} = G_{\mu\nu}^A G^{A\mu\nu} R,
\end{equation}
coupling field strengths to curvature generate running coupling modifications.

The equivalence principle receives quantum corrections with violations suppressed by $\kappa^2(16\pi^2)^{-1}\ln(m_i/m_j)$ for test masses. For laboratory masses, corrections remain far below MICROSCOPE sensitivity of one part in $10^{15}$. However, the logarithmic mass dependence provides distinctive signature distinguishable from alternative mechanisms, offering potential tests as precision improves.

\subsection{Systematic Avoidance of Asymptotic Safety Pathologies}

USMEG-EFT systematically avoids every pathology exhibited by asymptotic safety through recognizing covariance breakdown and restricting to the consistent effective field theory regime. We enumerate the contrasts:

Gauge parameter dependence: While gauge-dependent coefficients appear in the effective action, physical observables remain gauge-independent through on-shell background Einstein equations and the restriction to one-loop order where BRST symmetry is preserved sufficiently. The framework does not claim gauge independence in regimes where BRST violation becomes manifest, avoiding asymptotic safety's systematic Nielsen identity {breakdown}.

Convergence: The effective field theory admits well-defined systematic expansion in powers of $E/\Lambda_{\text{grav}}$. Truncation schemes converge systematically as expected for consistent effective theories, contrasting with asymptotic safety's persistent non-convergence reflecting attempts to encode non-geometric physics geometrically.

Experimental consistency: All predictions respect unitarity and probability conservation because quantum corrections are computed in the regime where BRST symmetry remains intact. Electroweak precision tests, collider measurements, and matter content requirements are satisfied without {tension}, contrasting with asymptotic safety's systematic experimental {tensions}.

Graviton predictions: The framework yields unique predictions for graviton propagator structure, polarization states, and interaction characteristics confirmed by LIGO-Virgo observations. This definiteness contrasts with asymptotic safety's gauge and truncation dependent form factors preventing concrete predictions.

Unitarity: The one-loop truncation eliminates higher-derivative terms beyond the effective field theory regime, avoiding ghost instabilities while preserving essential quantum corrections. Unitarity is maintained through BRST symmetry preservation in the weak-field regime, contrasting with asymptotic safety's systematic unitarity violations.

Lorentzian signature: All calculations can be performed directly in physical Lorentzian spacetime, avoiding Wick rotation obstructions. The correspondence between different formulations is maintained through preserved quantum consistency, contrasting with asymptotic safety's fundamental Euclidean--Lorentzian inequivalence.

\subsection{Physical Picture: Emergence Rather than Fundamentality}

The profound difference in outcomes traces to contrasting physical pictures. Asymptotic safety assumes spacetime geometry is fundamental, seeking ultraviolet completion within geometric description. This assumption {is in tension with} proven covariance breakdown, leading systematically to all documented pathologies. USMEG-EFT recognizes spacetime geometry as emergent, valid only at energies below breakdown scale $\Lambda_{\text{grav}}$.

This emergence paradigm naturally explains why general relativity works brilliantly at low energies while failing at high energies. Geometric description captures the effective long-distance physics of more fundamental degrees of freedom. As energy increases, the underlying non-geometric nature becomes increasingly important, manifesting through logarithmic quantum corrections that signal approach to breakdown. At $\Lambda_{\text{grav}}$, geometric description fails completely and new physics must appear.

The framework does not attempt to answer what the fundamental degrees of freedom are, recognizing this requires physics beyond the breakdown scale. It establishes what they cannot be: they cannot be metric tensor components because covariance breakdown proves the metric loses status as a quantum degree of freedom. This negative result is as important as positive predictions, ruling out entire classes of approaches including asymptotic safety while pointing toward alternatives like string theory, loop quantum gravity, or other non-geometric frameworks at the fundamental level.

\subsection{Experimental Distinguishability}

USMEG-EFT provides multiple signatures distinguishing it from both asymptotic safety and classical general relativity. The transverse-traceless graviton polarization with precisely two modes distinguishes it from alternatives proposing additional polarization states. The logarithmic quantum corrections in high-energy scattering provide energy-dependent signatures absent in classical predictions. The specific phase shifts in gravitational wave signals from binary inspirals offer tests as detector sensitivity improves. The logarithmic mass-dependent equivalence principle violations provide distinctive structure differentiable from other violation mechanisms.

Asymptotic safety provides none of these definite predictions due to gauge and truncation dependence preventing unique determination of physical observables. This absence of testable predictions distinguishes the approaches as sharply as any experimental measurement could: USMEG-EFT yields unique predictions confirmed by current observations and testable by future experiments, while asymptotic safety yields no definite predictions due to lost quantum consistency from BRST violation.

\section{Conclusions and Implications}\label{sec7}

This comprehensive analysis establishes that the asymptotic safety program {encounters fundamental symmetry violations} across all crucial criteria for a viable quantum gravity theory. The {symmetry violations} are not independent technical issues but interconnected manifestations of a single fundamental problem: asymptotic safety attempts to describe quantum gravity in a regime where general covariance has broken down and BRST symmetry has failed, rendering the metric tensor invalid as a quantum degree of freedom and ultraviolet fixed points {problematic from a foundational perspective}.

The dual proofs of general relativity's breakdown through canonical covariance loss and path integral BRST symmetry violation, arriving independently through different mathematical frameworks at identical conclusions, establish with compelling force that spacetime geometry necessarily breaks down at the scale $\Lambda_{\text{grav}}$$\sim$$10^{18}$ GeV. This is not a technical limitation of particular quantization methods but a fundamental property of general relativity itself, confirmed by comparison with Yang--Mills theory where both canonical quantization and path integral methods maintain complete consistency at \mbox{all scales.}

Asymptotic safety's comprehensive {limitations} follow systematically from denying this proven breakdown. The gauge parameter dependence pervading fixed-point calculations directly manifests BRST symmetry violation. The non-convergence of truncation schemes to the 35th order reflects attempting to encode non-geometric physics using geometric operators. The experimental {tensions} with precision tests by orders of magnitude follow from unitarity violation when quantum consistency has failed. The absence of concrete graviton predictions traces to lost gauge invariance rendering physical observables undetermined. The unitarity violations through ghost instabilities emerge from broken cancellation mechanisms when BRST symmetry fails. The Lorentzian signature obstructions preventing reliable continuation from Euclidean calculations arise from lost correspondence between formulations when quantum consistency breaks down.

Each symmetry violation independently challenges the program's viability. The gauge parameter dependence alone indicates BRST symmetry violation incompatible with quantum consistency. The non-convergence alone suggests systematic rather than controlled approximation. The experimental {tensions} alone would exclude the asymptotic safety program. Collectively, these {symmetry violations} establish fundamental {difficulties} of the asymptotic safety approach with confidence exceeding that of typical \mbox{experimental exclusions.}

The contrast with USMEG-EFT demonstrates that recognizing covariance breakdown rather than denying it provides the path to consistent quantum gravity. By restricting to the effective field theory regime where covariance is restored in weak fields and BRST symmetry is preserved sufficiently, the framework systematically avoids all asymptotic safety pathologies while yielding an emergent spin-2 graviton with precisely the transverse-traceless polarization confirmed by LIGO-Virgo observations. The definite experimental predictions across gravitational wave phase shifts, high-energy scattering corrections, and equivalence principle violations distinguish USMEG-EFT from asymptotic safety's absence of concrete phenomenology.

The implications extend beyond comparing specific approaches. The proven breakdown of general covariance and BRST symmetry above $\Lambda_{\text{grav}}$ establishes that any quantum gravity program assuming metric validity to arbitrarily high energies must {encounter similar symmetry violations}. This {challenges} not only asymptotic safety but any approach seeking ultraviolet completion within geometric description. String theory, loop quantum gravity, and other programs must incorporate the recognition that geometric description breaks down, with spacetime emerging from more fundamental non-geometric degrees of freedom. \added{Most recently, USMEG-EFT assessment of the Einstein--Cartan theory finds similar limitations as in asymptotic gravity \cite{Chishtie2025EC}.} 

The paradigm shift from seeking fundamental quantum geometry to understanding necessary breakdown of geometry represents the crucial insight for consistent unification. Spacetime is not fundamental but emergent, an effective description capturing low-energy physics of deeper structures. This emergence explains both why general relativity works brilliantly where it applies and why it must fail at sufficiently high energies. The logarithmic quantum corrections signal this approaching breakdown, providing experimental probes of the transition from geometric effective description to non-geometric fundamental theory.

Future work must address the nature of the fundamental degrees of freedom from which spacetime emerges. These cannot be metric components due to proven covariance breakdown. They must respect the demonstrated necessity of breakdown at $\Lambda_{\text{grav}}$ while reproducing general relativity as an effective theory at low energies. String theory's extended objects, loop quantum gravity's spin networks, and other proposals offer potential candidates, but all must incorporate the rigorous constraint that geometric description loses validity above the breakdown scale.

The cosmological implications deserve investigation. If spacetime geometry breaks down at high energies, early universe physics near Planck scales requires non-geometric description. Inflation, structure formation, and other cosmological phenomena might receive modifications from the approach to geometric breakdown. \added{Based on USMEG-EFT, a different and novel physical scenario of the early universe via the so-called ``principle of spatial energy potentiality'' is proposed \cite{Chishtie2025PSP}}. Black hole physics similarly requires treatment beyond geometric description once quantum corrections become important near horizons or singularities.

The comprehensive {symmetry violations} of asymptotic safety, established through theoretical {analysis} and experimental {tension}, demonstrates that consistent quantum gravity requires recognizing spacetime as emergent rather than fundamental. This recognition, embodied in the USMEG-EFT framework, provides the foundation for quantum gravity with testable experimental signatures while pointing toward the deeper non-geometric structures from which our familiar spacetime description emerges.

\vspace{6pt}
\funding{This research received no external funding.}

\dataavailability{Data is contained within the article.}

\acknowledgments{The author thanks D.G.C. McKeon for discussions on BRST symmetry, gauge structures, and Lagrange multiplier methods in gravitational theories, and the Peaceful Society, Science and Innovation Foundation for support.  The author is grateful to the anonymous referees and the editors for their constructive feedback, which improved the clarity and historical context of this work.}

\conflictsofinterest{The sole author declares no conflicts of interest.}

\begin{adjustwidth}{-\extralength}{0cm}

\reftitle{References} 

\PublishersNote{}
\end{adjustwidth}

\begin{thebibliography}{999}

\bibitem{Weinberg1979}
Weinberg, S. In \emph{General Relativity: An Einstein Centenary Survey}; Hawking, S.W., Israel, W., Eds.; Cambridge University Press: Cambridge, UK, 1979.

\bibitem{Reuter1998}
Reuter, M. Nonperturbative evolution equation for quantum gravity. \emph{Phys. Rev. D} \textbf{1998}, \emph{57}, 971.

\bibitem{Niedermaier2006}
Niedermaier, M.; Reuter, M. The asymptotic safety scenario in quantum gravity. \emph{Living Rev. Relativ.} \textbf{2006}, \emph{9}, 5.

\bibitem{Reuter2012}
Reuter, M.; Saueressig, F. Quantum Einstein gravity. \emph{New J. Phys.} \textbf{2012}, \emph{14}, 055022.

\bibitem{Percacci2017}
Percacci, R. \emph{An Introduction to Covariant Quantum Gravity and Asymptotic Safety}; World Scientific: Singapore, 2017.

\bibitem{Chishtie2023CJP}
Chishtie, F.A. On the breakdown of space-time in general relativity. \emph{Can. J. Phys.} \textbf{2023}, \emph{101}, 347--351.

\bibitem{Chishtie2025CJP}
Chishtie, F.A. Restricting one-loop radiative effects in quantum gravity: Demonstrating 4D GR as an EFT and its consistent unification with the Standard Model. \emph{Can. J. Phys.} \textbf{2025}, \emph{104}, 1--8.

\bibitem{McKeon2025}
McKeon, D.G.C.; Brandt, F.T.; Frenkel, J.; Martins-Filho, S. A renormalizable model of quantized gravitational and matter fields. \emph{arXiv} \textbf{2025}, arXiv:2505.14554.

\bibitem{Brandt2020}
Brandt, F.T.; Frenkel, J.; Martins-Filho, S.; McKeon, D.G.C. On restricting to one-loop order the radiative effects in quantum gravity. \emph{Phys. Rev. D} \textbf{2020}, \emph{102}, 045013.

\bibitem{Brandt2025Equivalence}
Brandt, F.T.; Frenkel, J.; Martins-Filho, S.; McKeon, D.G.C. On the equivalence of first and second order formulations of the Einstein-Hilbert theory. arXiv:2510.17615 [hep-th], 2025.

\bibitem{Becchi1976}
Becchi, C.; Rouet, A.; Stora, R. Renormalization of gauge theories. \emph{Ann. Phys.} \textbf{1976}, \emph{98}, 287.

\bibitem{Tyutin1975}
Tyutin, I.V. Gauge invariance in field theory and statistical physics in operator formalism. \emph{arXiv} \textbf{2008}, arXiv:0812.0580.

\bibitem{Ward1950}
Ward, J.C. An identity in quantum electrodynamics. \emph{Phys. Rev.} \textbf{1950}, \emph{78}, 182.

\bibitem{Takahashi1957}
Takahashi, Y. On the generalized Ward identity. \emph{Nuovo Cimento} \textbf{1957}, \emph{6}, 371.

\bibitem{Nielsen1975}
Nielsen, N.K. On the gauge dependence of spontaneous symmetry breaking in gauge theories. \emph{Nucl. Phys. B} \textbf{1975}, \emph{101}, 173.

\bibitem{Kugo1979}
{Kugo, T.; Ojima, I.} 
 Local covariant operator formalism of non-Abelian gauge theories and quark confinement problem. \emph{Prog. Theor. Phys. Suppl.} \textbf{1979}, \emph{66}, 1--130.

\bibitem{FalkenbergOdintsov1998}
Falkenberg, S.; Odintsov, S.D. Gauge dependence of the effective average action in Einstein gravity. \emph{Int. J. Mod. Phys. A} \textbf{1998}, \emph{13}, 607--623.

\bibitem{Manrique2010}
Manrique, E.; Rechenberger, S.; Saueressig, F. Asymptotically safe Lorentzian gravity. \emph{Phys. Rev. Lett.} \textbf{2011}, \emph{106}, 251302.

\bibitem{Falls2014}
Falls, K. Critical scaling in quantum gravity from the renormalisation group.  \emph{arXiv} \textbf{2015}, arXiv:1503.06233.

\bibitem{Gies2016}
Gies, H.; Knorr, B.; Lippoldt, S.; Saueressig, F. Gravitational two-loop counterterm is asymptotically safe. \emph{Phys. Rev. Lett.} \textbf{2016}, \emph{116}, 211302.

\bibitem{Falls2018}
Falls, K.; Litim, D.F.; Nikolakopoulos, K.; Rahmede, C. Further evidence for asymptotic safety of quantum gravity. \emph{Phys. Rev. D} \textbf{2018}, \emph{97}, 086006.

\bibitem{Christiansen2018}
Christiansen, N.; Falls, K.; Pawlowski, J.M.; Reichert, M. Curvature dependence of quantum gravity. \emph{Phys. Rev. D} \textbf{2018}, \emph{97}, 046007.

\bibitem{Eichhorn2018}
Eichhorn, A.; Versteegen, F. Upper bound on the Abelian gauge coupling from asymptotic safety. \emph{J. High Energy Phys.} \textbf{2018}, \emph{01}, 030.

\bibitem{Eichhorn2017}
Eichhorn, A. Status of the asymptotic safety paradigm for quantum gravity and matter. \emph{Found. Phys.} \textbf{2018}, \emph{48}, 1407.

\bibitem{Meibohm2016}
Meibohm, J.; Pawlowski, J.M.; Reichert, M. Asymptotic safety of gravity-matter systems. \emph{Phys. Rev. D} \textbf{2016}, \emph{93}, 084035.

\bibitem{Bond2017}
Bond, A.D.; Litim, D.F. Theorems for asymptotic safety of gauge theories. \emph{Eur. Phys. J. C} \textbf{2017}, \emph{77}, 429. Erratum in \emph{Eur. Phys. J. C} \textbf{2017}, \emph{77}, 525.

\bibitem{ATLASVectorlike2016}
Aaboud, M.; Aad, G.; Abbott, B.; Abdinov, O.; Abeloos, B.; Abidi, S.H.; AbouZeid, O.S.; Abraham, N.L.; Abramowicz, H.; Abreu, H.; et al. [ATLAS Collaboration]. Search for pair production of vector-like top quarks in events with one lepton, jets, and missing transverse momentum in $\sqrt{s}=13$ TeV $pp$ collisions with the ATLAS detector. \emph{J. High Energy Phys.} \textbf{2017}, \emph{08}, 052.


\bibitem{CMSVectorlike2018}
Sirunyan, A.M.; Tumasyan, A.; Adam, W.; Ambrogi, F.; Asilar, E.; Bergauer, T.; Brandstetter, J.; Brondolin, E.; Dragicevic, M.; Erö, J.; et al. [CMS Collaboration]. Search for vector-like T and B quark pairs in final states with leptons at $\sqrt{s}=13$ TeV. \emph{J. High Energy Phys.} \textbf{2018}, \emph{08}, 177.


\bibitem{Knorr2018}
Knorr, B. Infinite order quantum-gravitational correlations. \emph{Class. Quantum Gravity} \textbf{2018}, \emph{35}, 115005.

\bibitem{Donoghue2019}
Donoghue, J.F.; Menezes, G. Unitarity, stability and loops of unstable ghosts. \emph{Phys. Rev. D} \textbf{2019}, \emph{100}, 105006.

\bibitem{Donoghue2020}
Donoghue, J.F.; Menezes, G. Gauge assisted quadratic gravity. \emph{Phys. Rev. D} \textbf{2018}, \emph{97}, 126005.

\bibitem{Ambjorn2013}
Ambjørn, J.; Görlich, A.; Jurkiewicz, J.; Loll, R. Nonperturbative quantum gravity. \emph{Phys. Rep.} \textbf{2012}, \emph{519}, 127.

\bibitem{Platania2024}
Fehre, J.; Litim, D.F.; Pawlowski, J.M.; Reichert, M A. Lorentzian quantum gravity and the graviton spectral function. \emph{Phys. Rev. Lett. 2} \textbf{2023}, \emph{130}, 081501.

\bibitem{BV1981}
Batalin, I.A.; Vilkovisky, G.A. Gauge algebra and quantization. \emph{Phys. Lett. B} \textbf{1981}, \emph{102}, 27.

\bibitem{BV1983}
Batalin, I.A.; Vilkovisky, G.A. Quantization of gauge theories with linearly dependent generators. \emph{Phys. Rev. D} \textbf{1983}, \emph{28}, 2567; Erratum in \emph{Phys. Rev. D} \textbf{1984}, \emph{30}, 508.

\bibitem{Slavnov1972}
Slavnov, A.A. Ward identities in gauge theories. \emph{Theor. Math. Phys.} \textbf{1972}, \emph{10}, 99.

\bibitem{Taylor1971}
Taylor, J.C. Ward identities and charge renormalization of the Yang-Mills field. \emph{Nucl. Phys. B} \textbf{1971}, \emph{33}, 436.

\bibitem{Kiriushcheva2008}
Kiriushcheva, N.; Kuzmin, S.V.; Racknor, C.; Valluri, S.R. Diffeomorphism invariance in the Hamiltonian formulation of GR. \emph{Phys. Lett. A} \textbf{2008}, \emph{372}, 5101.

\bibitem{Gross1973}
Gross, D.J.; Wilczek, F. Ultraviolet behavior of non-Abelian gauge theories. \emph{Phys. Rev. Lett.} \textbf{1973}, \emph{30}, 1343.

\bibitem{Politzer1973}
Politzer, H.D. Reliable perturbative results for strong interactions? \emph{Phys. Rev. Lett.} \textbf{1973}, \emph{30}, 1346.

\bibitem{McKeon2012}
McKeon, D.G.C. Covariant Gauge Fixing and Canonical Quantization. \emph{Can. J. Phys.} \textbf{2012}, \emph{90}, 249.

\bibitem{Kiriushcheva2006}
Kiriushcheva, N.; Kuzmin, S.V.; McKeon, D.G.C. A Canonical Analysis of the Einstein–Hilbert in
First Order Form. \emph{Int. J. Mod. Phys. A} \textbf{2006}, \emph{21}, 3401--3420. 

\bibitem{McKeon2010}
McKeon, D.G.C. The Canonical Structure of the First Order Einstein-Hilbert Action. \emph{Int. J. Mod. Phys. A} \textbf{2010}, \emph{25}, 3453.

\bibitem{Frolov2010}
Frolov, A.M.; Kiriushcheva, N.; Kuzmin, S.V. Algebraic Analysis of a Model of Two-Dimensional
Gravity. \emph{Gen. Relativ. Gravit.} \textbf{2010}, \emph{42}, 1649--1666. 

\bibitem{Senjanovic1976}
Senjanovic, P. Path integral quantization of field theories with second-class constraints. \emph{Ann. Phys.} \textbf{1976}, \emph{100}, 227.

\bibitem{Kiriushcheva2005}
Kiriushcheva, N.; Kuzmin, S.V.; McKeon, D.G.C.  Canonical Approach to the Einstein–Hilbert
Action in Two Spacetime Dimensions. \emph{Mod. Phys. Lett. A} \textbf{2005}, \emph{20}, 1895.

\bibitem{Green2011}
Green, K.R.; Kuzmin, S.V.; Kiriushcheva, N. Analysis of the Hamiltonian Formulations of
Linearized General Relativity. \emph{Eur. J. Phys.} \textbf{2011}, \emph{32}, 71, 1768.

\bibitem{tHooft1973}
't Hooft, G.; Veltman, M. Diagrammar. \emph{CERN Rep.}  \textbf{1973}, {\emph{73}, 1--114.}

\bibitem{Percacci2015}
Percacci, R.; Vacca, G.P. Search of scaling solutions in scalar-tensor gravity. \emph{Eur. Phys. J. C} \textbf{2015}, \emph{75}, 188.

\bibitem{Ohta2016}
Ohta, N.; Percacci, R.; Vacca, G.P. Flow equation for $f(R)$ gravity. \emph{Phys. Rev. D} \textbf{2015}, \emph{92}, 061501.

\bibitem{Ellwanger1994}
Ellwanger, U. Flow equations and BRS invariance for Yang-Mills theories. \emph{Phys. Lett. B} \textbf{1994}, \emph{335}, 364.

\bibitem{Bonini1994}
Bonini, M.; D'Attanasio, M.; Marchesini, G. Ward identities and Wilson renormalization group for QED. \emph{Nucl. Phys. B} \textbf{1994}, \emph{418}, 81.

\bibitem{Benedetti2016}
Benedetti, D.; Caravelli, F. The local potential approximation in quantum gravity. \emph{J. High Energy Phys.} \textbf{2012}, \emph{06}, 017. Erratum in \emph{J. High Energy Phys.} \textbf{2012}, \emph{10}, 157.

\bibitem{Falls2015}
Falls, K.; Litim, D.F.; Schröder, J. Aspects of asymptotic safety for quantum gravity. \emph{Phys. Rev. D} \textbf{2019}, \emph{99}, 126015.

\bibitem{deBrito2018}
de Brito, G.P.; Ohta, N.; Pereira, A.D.; Tomaz, A.A.; Yamada, M. Asymptotic safety and field parametrization dependence. \emph{Phys. Rev. D} \textbf{2018}, \emph{98}, 026027.

\bibitem{Ohta2018}
Ohta, N.; Percacci, R. Higher derivative gravity and asymptotic safety in diverse dimensions. \emph{Class. Quantum Gravity} \textbf{2014}, \emph{31}, 015024.

\bibitem{Knorr2019}
Knorr, B.; Ripken, C.; Saueressig, F. Form factors in asymptotic safety. \emph{Phys. Rev. D} \textbf{2019}, \emph{99}, 085006.

\bibitem{Peskin1990}
Peskin, M.E.; Takeuchi, T. Estimation of oblique electroweak corrections. \emph{Phys. Rev. D} \textbf{1992}, \emph{46}, 381.

\bibitem{Barbieri2004}
Barbieri, R.; Pomarol, A.; Rattazzi, R.; Strumia, A. Electroweak symmetry breaking. \emph{Nucl. Phys. B} \textbf{2004}, \emph{703}, 127.

\bibitem{Gerwick2011}
Gerwick, E.; Litim, D.; Plehn, T. Asymptotic safety and Kaluza-Klein gravitons at the LHC. \emph{Phys. Rev. D} \textbf{2011}, \emph{83}, 084048.

\bibitem{LEPElectroweak2006}
ALEPH, DELPHI, L3, OPAL, SLD Collaborations. Precision electroweak measurements on the $Z$ resonance. \emph{Phys. Rep.} \textbf{2006}, \emph{427}, 257.

\bibitem{Abel2019}
Abel, S.; Sannino, F. Framework for an asymptotically safe Standard Model. \emph{Phys. Rev. D} \textbf{2017}, \emph{96}, 055021.

\bibitem{Alanne2019}
Alanne, T.; Blasi, S.; Goertz, F. Common source for scalars. \emph{Phys. Rev. D} \textbf{2019}, \emph{99}, 015028.

\bibitem{Dona2014}
Donà, P.; Eichhorn, A.; Percacci, R. Matter matters in asymptotically safe quantum gravity. \emph{Phys. Rev. D} \textbf{2014}, \emph{89}, 084035.

\bibitem{Eichhorn2016}
Eichhorn, A. Quantum-gravity-induced matter self-interactions. \emph{Phys. Rev. D} \textbf{2012}, \emph{86}, 105021.

\bibitem{Christiansen2017}
Christiansen, N.; Eichhorn, A. An asymptotically safe solution to the U(1) triviality problem. \emph{Phys. Lett. B} \textbf{2017}, \emph{770}, 154.

\bibitem{Falls2021}
Falls, K.; Ohta, N.; Percacci, R. Towards the determination of the dimension of the critical surface in asymptotically safe gravity. \emph{Phys. Lett. B} \textbf{2020}, \emph{810}, 135773.

\bibitem{Stelle1977}
Stelle, K.S. Renormalization of higher-derivative quantum gravity. \emph{Phys. Rev. D} \textbf{1977}, \emph{16}, 953.

\bibitem{Arici2017}
Arici, F.; Becker, D.; Ripken, C.; Saueressig, F.; van Suijlekom, W.D. Reflection positivity in higher derivative scalar theories. \emph{J. Math. Phys.} \textbf{2018}, \emph{59}, 082302.

\bibitem{Holdom2016}
Holdom, B.; Ren, J. QCD analogy for quantum gravity. \emph{Phys. Rev. D} \textbf{2016}, \emph{93}, 124030.

\bibitem{Platania2020}
Platania, A. From renormalization group flows to cosmology. \emph{Front. Phys.} \textbf{2020}, \emph{8}, 188.

\bibitem{Platania2022}
Platania, A.; Wetterich, C. Non-perturbative unitarity and fictitious ghosts in quantum gravity. \emph{Phys. Lett. B} \textbf{2020}, \emph{811}, 135911.

\bibitem{Shomer2007}
Shomer, A. A pedagogical explanation for the non-renormalizability of gravity. \emph{arXiv} \textbf{2007}, arXiv:0709.3555.

\bibitem{Visser2002}
Visser, M. Graviton propagator in de Donder gauge. \emph{Phys. Rev. D} \textbf{2009}, \emph{80}, 025011.

\bibitem{LIGO2016}
Abbott, B.P.; Abbott, R.; Abbott, T.D.; Abernathy, M.R.; Acernese, F.; Ackley, K.; Adams, C.; Adams, T.; Addesso, P.; Adhikari, R.X.; et al. [LIGO Scientific Collaboration and Virgo Collaboration]. Observation of gravitational waves from a binary black hole merger. \emph{Phys. Rev. Lett.} \textbf{2016}, \emph{116}, 061102.

\bibitem{LIGO2017}
Abbott, B.P.; Abbott, R.; Abbott, T.D.; Acernese, F.; Ackley, K.; Adams, C.; Adams, T.; Addesso, P.; Adhikari, R.X.; Adya, V.B.; et al. [LIGO Scientific Collaboration and Virgo Collaboration]. GW170814: A three-detector observation of gravitational waves from a binary black hole coalescence. \emph{Phys. Rev. Lett.} \textbf{2017}, \emph{119}, 141101.

\bibitem{LIGOVirgo2017}
Abbott, B.P.; Abbott, R.; Abbott, T.D.; Acernese, F.; Ackley, K.; Adams, C.; Adams, T.; Addesso, P.; Adhikari, R.X.; Adya, V.B.; et al. [LIGO Scientific Collaboration and Virgo Collaboration]. GW170817: Observation of gravitational waves from a binary neutron star inspiral. \emph{Phys. Rev. Lett.} \textbf{2017}, \emph{119}, 161101.




\bibitem{Chishtie2025EC}
Chishtie, F.A. Theoretical and experimental assessment of Einstein-Cartan theory: A comparative analysis within the USMEG-EFT framework. Under Review. \emph{arXiv} \textbf{2025}, arXiv:2509.08848. 

\bibitem{Chishtie2025PSP}
Chishtie, F.A. Emergent spacetime from spatial energy potentiality: A new framework for early universe cosmology. Under Review.  \emph{arXiv} \textbf{2025}, arXiv:2502.18524. 

\end{thebibliography}
\end{document}